\documentclass[reprint,amsmath,amssymb,pra,aps,nofootinbib,superscriptaddress,floatfix]{revtex4-1}

\usepackage[T1]{fontenc}
\usepackage[utf8]{inputenc}
\usepackage{amsmath, amssymb}
\usepackage{bbold}
\usepackage{bm}
\usepackage{tipa}
\usepackage{subfigure}
\usepackage{blindtext}
\usepackage[dvipsnames]{xcolor}
\definecolor{dcyan}{RGB}{0,100,100}
\usepackage{graphicx}
\usepackage{tabularx}
\usepackage{textgreek}
\usepackage{xfrac}
\usepackage{units}
\usepackage{comment}
\usepackage{float}
\usepackage{booktabs}
\usepackage{makecell}
\usepackage{array}
\usepackage{enumitem}
\usepackage{upgreek}
\usepackage{hyperref}
\setlist[enumerate]{topsep=2pt,itemsep=1pt,parsep=0pt,partopsep=0pt}
\setlist[enumerate,2]{topsep=1pt,itemsep=0.5pt}

\usepackage{xcolor}
\definecolor{green_cust}{RGB}{0,154,85}
\definecolor{red_cust}{RGB}{173,49,54}
\definecolor{blue_cust}{RGB}{0,103,148}
\hypersetup{colorlinks=true, linkcolor=red_cust, citecolor=green_cust, filecolor=blue_cust, urlcolor=blue_cust}

\makeatletter 
    
\renewcommand\onecolumngrid{
\do@columngrid{one}{\@ne}%
\def\set@footnotewidth{\onecolumngrid}
\def\footnoterule{\kern-6pt\hrule width 1.5in\kern6pt}%
}

\renewcommand\twocolumngrid{
        \def\footnoterule{
        \dimen@\skip\footins\divide\dimen@\thr@@
        \kern-\dimen@\hrule width.5in\kern\dimen@}
        \do@columngrid{mlt}{\tw@}
}%

\makeatother    

\usepackage{soul}

\newcommand{\Figref}[1]{Fig.~\hyperref[#1]{\ref{#1}}}

\newcommand{\Applied}{Department of Applied Physics, Stanford University, Stanford, CA}
\newcommand{\Physics}{Department of Physics, Stanford University, Stanford, CA}
\newcommand{\Chicago}{Department of Physics, The University of Chicago, Chicago, IL}

\begin{document}
\title{Stability, degeneracy, and scalability of a 600-site cavity array microscope}

\author{Anna Soper}\thanks{These authors contributed equally.}
\affiliation{\Applied}
\author{Danial Shadmany}\thanks{These authors contributed equally.}
\affiliation{\Physics}
\author{Adam L. Shaw}
\affiliation{\Applied}
\affiliation{\Physics}
\author{Lukas Palm}
\affiliation{\Chicago}
\author{David I. Schuster}
\affiliation{\Applied}
\author{Jonathan Simon}\email{jonsimon@stanford.edu}
\affiliation{\Applied}
\affiliation{\Physics}

\date{\today}


\begin{abstract}
Optical cavities are a foundational technology for controlling light-matter interactions. While interfacing a single cavity to either an atom or ensemble has become a standard tool, the advent of single atom control in large atomic arrays has spurred interest in a new frontier of ``many-cavity QED,'' featuring many independent resonators capable of separately addressing individual quantum emitters. In this fast-evolving landscape, the \textit{cavity array microscope} was recently introduced\textemdash employing free space intra-cavity optics to engineer a two-dimensional array of tightly spaced cavity TEM$_{00}$ modes with wavelength-scale waists, ideally suited for interfacing with atom arrays. Here we realize the next-generation of this architecture, achieving hundreds of degenerate cavity modes with improved, uniform finesse, and explore the technical features of the system which will enable further scalability. In particular, we study imperfections, including optical aberrations, field of view constraints, array non-degeneracies, and losses from optical elements. We identify the sensitivity to these various vectors and exposit the control knobs and techniques necessary to align and operate the system in a stable manner. Ultimately, we lay out a pathway towards operation with tens of thousands of independent cavities while maintaining compatibility with existing atom arrays, paving the way to myriad applications including highly parallelized remote entanglement generation, fast and non-destructive mid-circuit readout, and the implementation of hybrid atom-photon Hamiltonians.

\end{abstract}
\maketitle

\section{Introduction}


Cavity quantum electrodynamics (QED) is the study and control of light-matter interactions when photons are trapped in a high-Q resonator~\cite{walther2006cavity}. The cavity-enhanced density of states allows for pristine isolation of single or multiple photon modes, which has been enabling in fields ranging from bio-imaging and spectroscopy~\cite{reynolds2025quantum,needham2024labelfree, kohler2021tracking}, to detection of gravitational waves~\cite{abbott2016observation}, and quantum metrology~\cite{harouche2001manipulating,leroux2010implementation, cooper2024graphstates}. To advance these varied areas, cavity engineering has expanded continuously with the development of ultra-high finesse optics~\cite{ding2025highfinesse,hood2001characterization}, nanophotonic~\cite{samutpraphoot2020strong} and fiber~\cite{hunger2010fiber} interfaces, introduction of intra-cavity lenses~\cite{juffmann2016multi, jaffe2022backscatter,shadmany2025cavity}, nonlinear optics~\cite{taneja2025lightcontrolled}, and integration with high-NA imaging systems~\cite{orsi2024cavity}. Despite these evolutions, an elusive challenge is that an optical cavity typically acts as a very high performance `single pixel', lacking spatial resolution across its mode when probing a set of emitters. 

To address this limitation, a new optical cavity design was recently debuted, the \textit{cavity array microscope} (CAM)~\cite{shaw2025cavarray}. This architecture employs intra-cavity lenses to create a two-dimensional array of many independent, small waist, tightly-spaced TEM$_{00}$ cavity modes, effectively realizing a new regime of many-cavity QED. The CAM can be applied in any context in which a `many-pixel' cavity interface is desirable, and is especially useful when interfacing optical cavities with large arrays of individually trapped atoms~\cite{deist2022mid,grinkemeyer2025error,hu2025site,hartung2024quantumnetwork,liu2023realization,dhordjevic2021entanglement}, where the parallelized atom-cavity interface can improve the rate of transducing quantum information between atoms and photons as part of, for example, a distributed quantum computer~\cite{sinclair2025faulttolerant}, or enable studies of new light-matter coupled Hamiltonians~\cite{makin2008quantum,hartmann2008quantum,greentree2006quantum}. This was the context for the first CAM demonstration~\cite{shaw2025cavarray}, where an array of single atoms was simultaneously interfaced with 43 independent cavities. This first generation exhibited several limitations: bulky optics were required $\approx1$~mm away from atoms limiting optical access and hosting surface charges; the number of accessible cavities was constrained; each cavity mode averaged over two `pixels'; and cavity finesse was low due to several non-ideal optical components.

\begin{figure*}[ht!] 
    \includegraphics{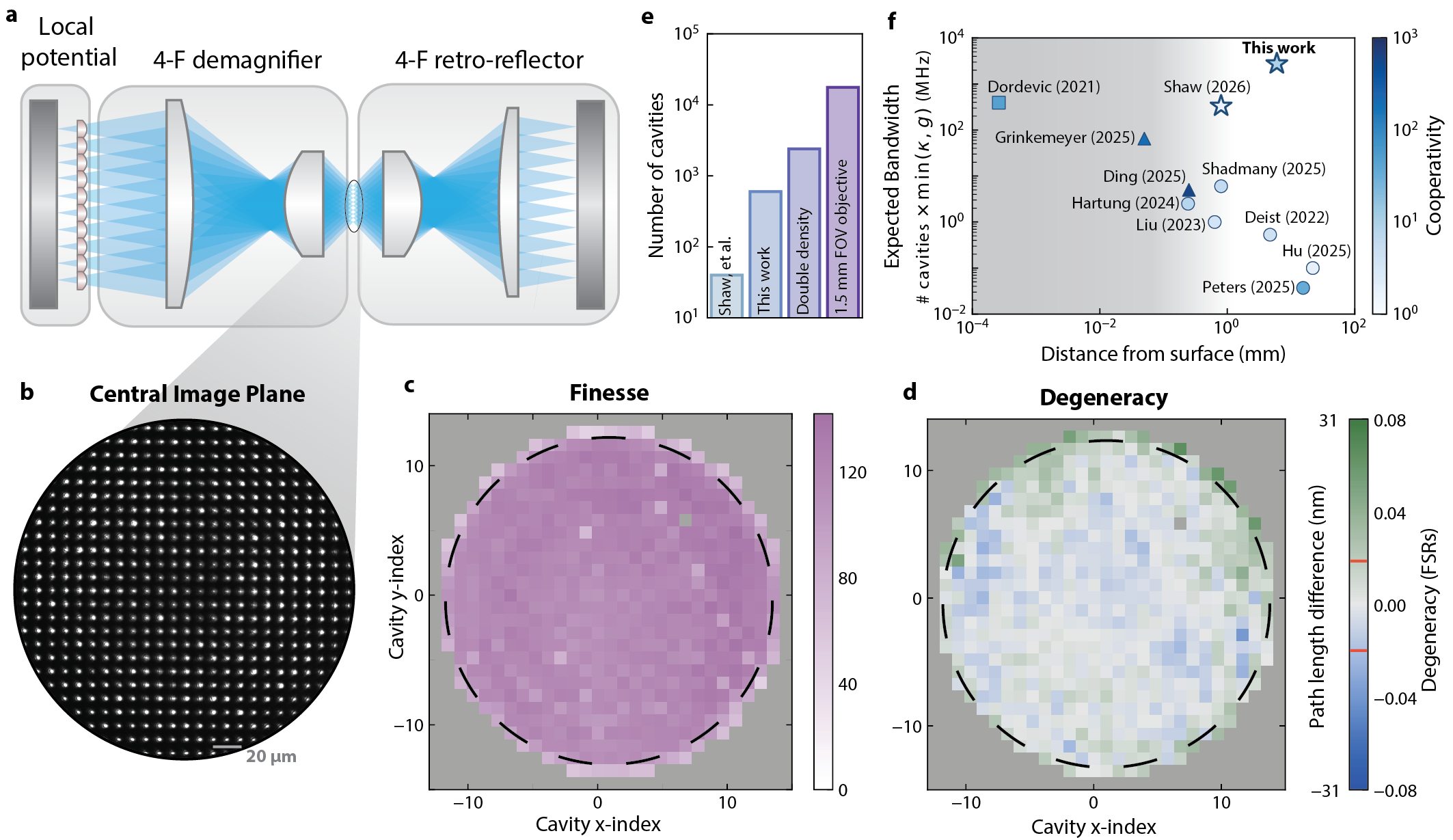}
    \caption{
        \textbf{Summary of results.} \textbf{a.} The cavity array microscope (CAM) is an optical imaging system that generates a 2D array of independent, simultaneously resonant optical cavities with wavelength-scale mode waists and mode spacings. \textbf{b.} Cavity transmission for hundreds of simultaneously driven individual cavity modes. \textbf{c, d.} We demonstrate a cavity array microscope with >600 cavities and an array-averaged finesse of 114(17). When the outcoupling is optimized for collection efficiency (Sup.~\ref{SI:collection}), 537 cavities are degenerate to within the cavity linewidth, indicated by red lines on color bar. The dashed black circle indicates the field of view of the intra-cavity optics, beyond which field curvature inhibits stability (Sec.~\ref{sec:fov}). The degeneracy of the CAM is interferometrically sensitive to nanometer-scale variations in optical path lengths between cavities due to misalignments, aberrations, stresses, and surface irregularity of various optics. \textbf{e.} The number of achievable cavities can be increased through simple modifications to the intra-cavity optics. By doubling the density of the microlens array (MLA) and leveraging a wider field of view microscope objective, the CAM could support tens of thousands of cavities at high cooperativity. \textbf{f.} We compare the CAM to other state-of-the-art, neutral atom array-compatible resonator geometries on the axes of bandwidth (fastest possible readout/entanglement speed), distance of atoms from dielectric surfaces, and cooperativity. Grey shading is a guide to the eye for distances at which surface charges are potentially deleterious to Rydberg excitation~\cite{ocola2025control}. Stars: cavity array microscopes~\cite{shaw2025cavarray}, Squares: nanophotonic cavities~\cite{dhordjevic2021entanglement}, Triangles: fiber and micro-mirror cavities~\cite{grinkemeyer2025error,ding2025highfinesse}, Circles: macro-mirror cavities~\cite{shadmany2025cavity,deist2022mid,hu2025site, liu2023realization, hartung2024quantumnetwork}.}
	\label{fig:summary}
\end{figure*}

In this work, we present a substantial evolution of the CAM design that overcomes these challenges, creating a scalable, high-performance cavity array interface with over 600 sites (Fig.~\ref{fig:summary}a,b). We anticipate that this improved architecture will establish the cavity array microscope as a versatile tool for a broad range of fields, from quantum computing and metrology to biological and chemical sensing (Sup.~\ref{SI:biosensing}). The key performance metrics for the system are its finesse (which at fixed waist determines the cavity cooperativity~\cite{tanji2011interaction}), the field of view (which determines the number of cavities), and the mutual degeneracy between cavities (which determines whether they can be simultaneously probed and read out). In this new realization each of these metrics are improved by an order of magnitude. Besides these direct quantitative improvements, we uncover the sources of error and aberration which limit each performance metric, and elucidate clear pathways to eliminate them.

Specifically, we obtain an average finesse of 114(17) across 603 cavities (Fig.~\ref{fig:summary}c), limited by coating losses and surface roughness of various optics. Paired with our nominal waist of $1.15~\upmu$m, this corresponds to an array-averaged single atom peak cooperativity $C>10$. The field of view of the CAM is 140~$\upmu$m (radius), determined by the numerical aperture (NA) required to achieve the expected waist and limited primarily by field curvature produced by the intra-cavity aspheric lens. Lastly, we achieve 537 cavities mutually degenerate within the readout-optimized cavity linewidth, wherein the transmission of the outcoupling mirror is increased, at the expense of cooperativity, to maximize the total efficiency of photons collected and outcoupled through the cavity (Fig.~\ref{fig:summary}d, Sup.~\ref{SI:collection}). The residual non-degeneracy is due to surface irregularity and stress on optics that results in nanometer-scale optical path length variations, which we identify through novel optical characterization techniques enabled by the CAM. With straightforward optics replacements, we project that the CAM can be scaled to tens of thousands of cavities (Fig.~\ref{fig:summary}e)\textemdash on par with the largest neutral atom arrays~\cite{chiu2025continuous,Lin2025AI,manetsch2024tweezer}.

Perhaps the biggest strength of our cavity design is its inherent scalability; both in the raw number of sites, as well as in our novel scheme for tuning all cavities to degeneracy through control over just a few degrees of freedom. 
Due to this high degree of parallelization, we envision the cavity array microscope as an ideal platform for quantum networking. We compare the projected performance of our system in terms of information bandwidth (defined as the number of cavities multiplied by the minimum of the cavity linewidth and cavity coupling, which limit the atom-photon entanglement rate) 
against other recent optical cavity designs (Fig.~\ref{fig:summary}f), predicting a state-of-the-art bandwidth while maintaining high cooperativity and keeping atoms far from dielectric surfaces. This latter point is important, as the short lengths, and thus small atom-surface separation of typical highly performant cavities, increase susceptibility to surface charges and thus decoherence of Rydberg atom operations~\cite{ocola2025control}. In contrast, our long working distance aspheric lenses realize a large 5.7~mm atom-surface separation. Thus, we expect the cavity array microscope to provide a viable path towards GHz-scale entangling rates, with the full capabilities of neutral atom quantum processors.

In the following sections we detail the capabilities and limitations of the cavity array microscope: in Section~\ref{sec:cav_design} we describe the experimental setup, paraxial cavity array stability, and design considerations; in Section~\ref{sec:fov} we combine data with numerics to understand the impact of field curvature on the cavity field of view; in Section~\ref{sec:finesse} we catalog the sources of loss that impact the cavity finesse; finally, in Section~\ref{sec:degeneracy} we identify sources of non-degeneracy, assess sensitivity to misalignment, and project achievable performance.

\section{Cavity Array Design}
Because the cavity array microscope (CAM) is constructed with multiple intra-cavity optics, it offers a highly configurable design space that can be customized to suit a multitude of applications, with flexible control of resonator length, cavity mode spacing, and mode waist. In this realization we employ a combination of off-the-shelf and custom optics to construct a CAM optimized to interface with a 2D neutral atom array in its central image plane. Here, we detail both the experimental parameters of this CAM, as well as the heuristics for customizing it to a generic set of design constraints. See Table~\ref{etab:params} for a summary of cavity parameters.

\label{sec:cav_design}
\subsection{Experimental setup}

\begin{figure*}[ht!] 
    \includegraphics{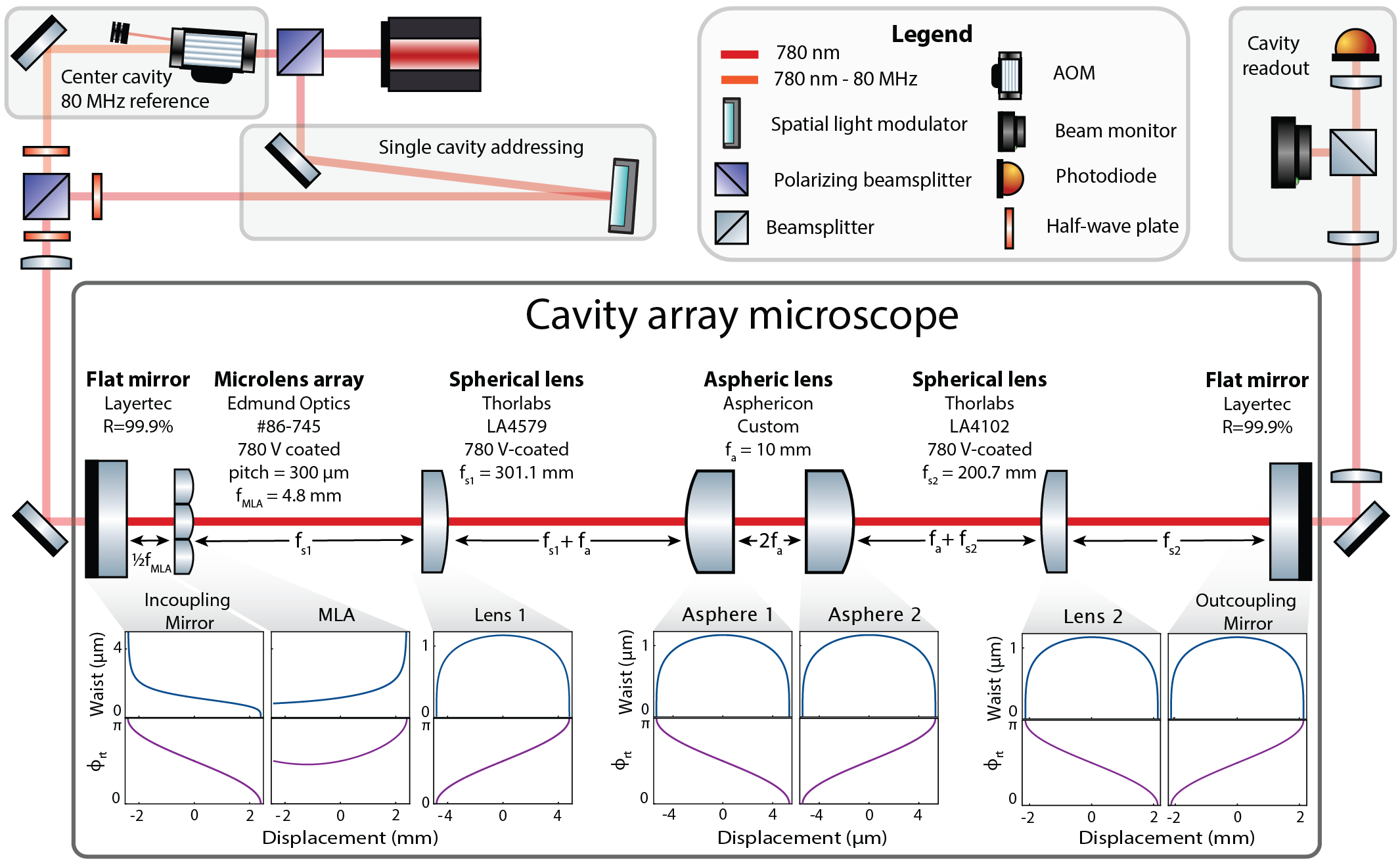}
    \caption{\textbf{Experimental setup.} Our experimental realization of the cavity array microscope (CAM) contains both off-the-shelf and custom optical elements. A spatial light modulator (SLM) is used to couple serially into each cavity in the array. The central cavity can be separately driven using light detuned by 80 MHz relative to the SLM path, enabling characterization of the relative degeneracy of different cavities to less than the cavity linewidth. All cavities are read out using a shared avalanche photodiode. \textbf{Lower panels:} Paraxial stability diagrams for the central cavity, plotting the smallest waist and round-trip Gouy phase ($\phi_{rt}$) as a function of displacements of each optic from its optimal telescopic position (indicated in the cavity schematic). The stability of all cavities in the array can be approximated by a paraxial model of the central ``on-axis'' cavity. The aspheric lens is the most sensitive optic, with its stability region spanning only 11~$\upmu$m of longitudinal displacement.
    }
	\label{fig:exp_setup}
\end{figure*}

The CAM has three primary components: 1) the microlens array (MLA), which acts as an array of optical potential wells, each defining a local ray-axis that returns to itself under a round trip through the cavity, and about which a family of paraxial TEM cavity modes can be expanded~\cite{sommer2016engineering}, 2) a primary demagnifying telescope that images the plane of the MLA into the central image plane where we plan to couple to an array of single atoms, and 3) a secondary retro-reflecting telescope that re-images the central image plane back to itself, thus avoiding the need for a mirror in the plane where the atoms will ultimately reside (Fig.~\ref{fig:summary}a). For an ideal set of lenses, this setup creates an array of cavities with identical mode waists and transverse mode spectra, where the waists are spatially separated yet tightly spaced in the central image plane (Fig.~\ref{fig:summary}b).

The principal design criteria for the CAM are the size and spacing of the mode waists. The spacing between mode waists is intuitively the demagnified MLA pitch, while the waist size is set by an effective ``short'' cavity consisting of the MLA and subsequent end-mirror (Fig.~\ref{fig:waist_heuristics}a). The waist size thus depends upon the MLA focal length, $f_{\textrm{MLA}}$, the demagnification, $M$, of the primary telescope, and the MLA-end mirror distance, $\Delta d_{\textrm{Mirror}}$. 
When the distance between the mirror and MLA is exactly $\Delta d_{\textrm{Mirror}}=f_{\textrm{MLA}}/2$, the mode waist in the central image plane can be written analytically as 

\begin{equation} \label{eq:1}
w_0=\frac{1}{M}\sqrt{\frac{\lambda\times f_{\textrm{MLA}}}{\pi}}
\end{equation}
(derivation detailed in the following subsection). 
We can understand this expression as the intrinsic waist set by the chosen focal length of the MLA, with an additional demagnification factor.


For our CAM we use an MLA with a 4.8~mm focal length and 300$~\upmu$m pitch, which sets an intrinsic waist of $\sqrt{\lambda \times f_{\textrm{MLA}}/\pi}=35$~$\upmu$m. We choose this short MLA focal length in order to reduce the required telescope magnification. Next, we use a 30$\times$ demagnifying primary telescope to shrink the mode waist to 1.15~$\upmu$m and the mode spacing to 10~$\upmu$m. For this telescope we employ a 300~mm focal length spherical lens paired with a 10~mm focal length high-NA aspheric lens~\cite{aspheric_lens} with a relatively large, 5.7~mm working distance. Finally, we complete the cavity with a 20$\times$ retro-reflecting secondary telescope composed of the same high-NA aspheric lens and a 200~mm spherical lens (Fig.~\ref{fig:exp_setup}). While the magnification of the primary telescope is chosen in conjunction with $f_{MLA}$ to set the cavity mode waist, the secondary telescope is unconstrained\textemdash without a microlens array on that side, the paraxial round trip ABCD matrix of the secondary telescope is just the identity and consequently has no impact on the mode waist of the central image plane. At 20$\times$, the magnified side of the telescope is relatively low NA, which allows us to use low loss spherical optics without significant impacts from spherical aberration (Sup.~\ref{SI:SphericalAberr}).

To better understand the sensitivity of the mode waist to misalignment of various optics, we create a paraxial model describing the central cavity of the CAM. Using this model, we compute the mode waist and Gouy phase with respect to longitudinal displacements of each optic from its ideal telescopic position (Fig.~\ref{fig:exp_setup}, lower panels) and find that the waist is either quadratically insensitive to perturbations (for most optics) or minimally sensitive to perturbations over millimeter length scales (for the incoupling mirror and the MLA). These calculations also reveal a narrow stability region for the aspheric lens\textemdash nearly three orders of magnitude smaller than that of any other optic. This means that the CAM's constituent cavities are particularly sensitive to $\upmu$m-scale drifts in the asphere axial positions, as well as optical aberrations that cause $\upmu$m-scale distortions to the asphere's focal plane (Sec.~\ref{sec:fov}). In practice, we find negligible axial drifts over week-long time scales (Sup.~\ref{SI:stability_time}).

In order to measure the performance of the CAM, we sequentially excite the cavities using a spatial light modulator (SLM) and read out the resulting spectra with a shared avalanche photodiode. For finesse measurements, the cavities are excited using a narrow ($<1$~kHz) linewidth, 780~nm laser, while a piezo sweeps the incoupling mirror through multiple cavity FSRs. Because the motion of the piezo interferes with the measurement of the CAM degeneracy, the degeneracy measurements are instead performed with a laser whose 
frequency is swept across many cavity FSRs. For the degeneracy measurements, we also add a 
beam permanently coupled into the central cavity that is shifted by 80~MHz relative to the SLM beam, serving as a static frequency reference (further details in Section~\ref{sec:degeneracy}).

\begin{table}[ht!]
    \centering
    \caption{\textbf{Cavity parameters.} Typical parameters of the cavity array microscope. Values in parentheses denote the standard deviation of array-averaged quantities. The mode waist is extracted from paraxial ABCD matrix calculations. Direct measurements of comparable $\lambda$-scale waists in related resonator geometries have been shown to agree with paraxial ABCD-matrix predictions within experimental uncertainty~\cite{shadmany2025cavity, shaw2025cavarray}.}
    \vspace{2mm}
    \begin{tabular}{l l c}
        \toprule
        \textbf{Mean parameter} & \textbf{\makecell{Value \ \ }} & \textbf{\makecell{Unit\\}} \\
        \midrule
        FSR   & 143.8 & MHz\\
        $g$ -- for ${}^{87}\text{Rb}$ cycling transition & $2\pi\times$4.4 & MHz\\
        $\kappa$ -- array averaged   & $2\pi\times$1.26 & MHz \\
        Finesse -- max & 145 & \\
        Finesse -- array averaged & 114(17) & \\
        Waist -- central focal plane & 1.15 & $\upmu$m \\
        Pitch -- central focal plane & 10 & $\upmu$m \\
        Length   & 1.0424  \ \ \ & m\\
        Achieved \# cavities & 603 & \\
        Geometric cooperativity -- array averaged \ \ \ \ & 10.2 & \\
        Projected collection efficiency (Sup.~\ref{SI:collection}) & 53 & \%\\
        \bottomrule
    \end{tabular}
    \label{etab:params}
\end{table}

\subsection{Design heuristics}
\begin{figure}[t!] 
    \centering
    \includegraphics[width=\columnwidth]{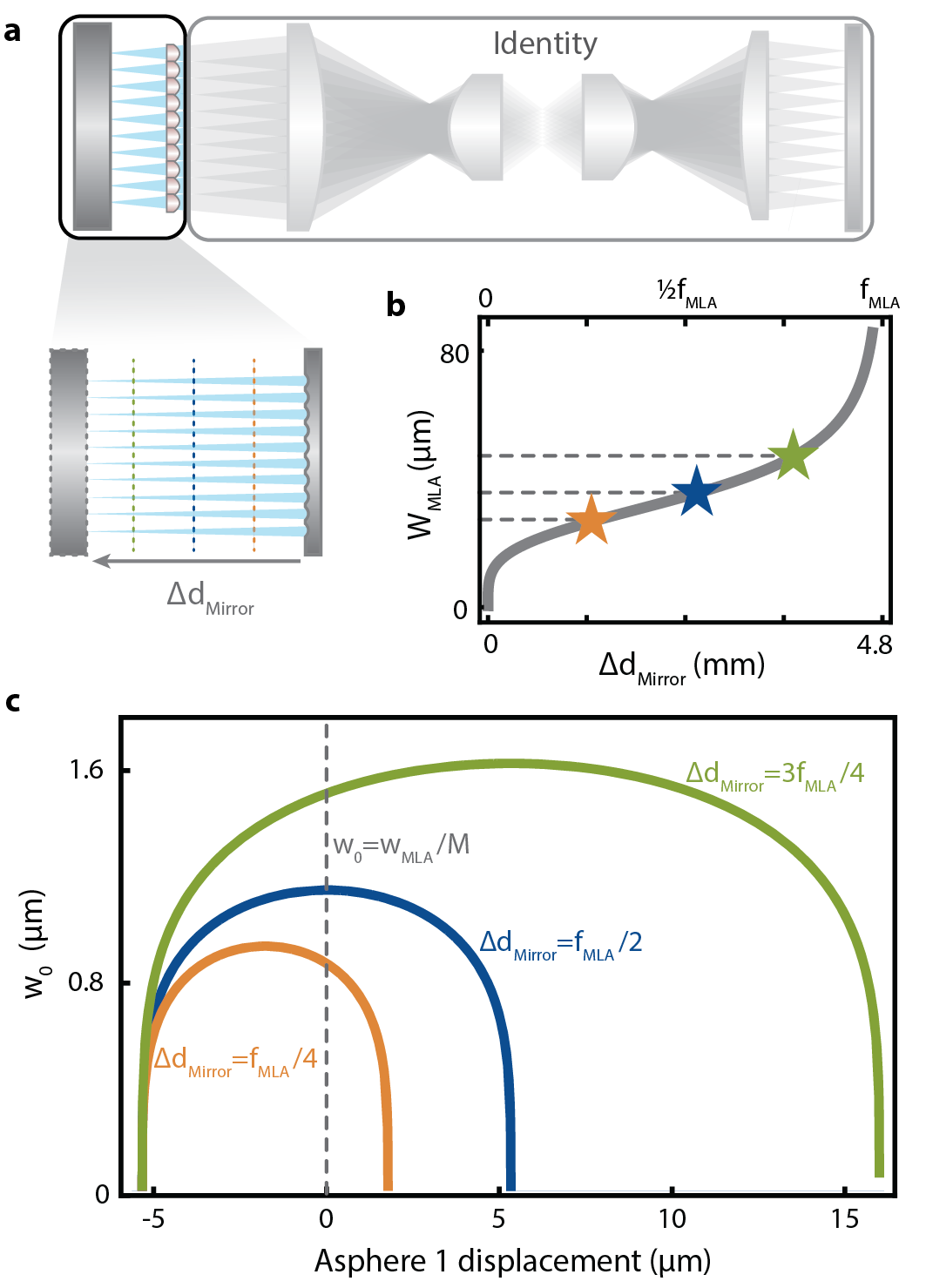}
    \caption{\textbf{Cavity mode waist.} \textbf{a.} The two 4f telescopes of the CAM have the action of the identity on an incoming ray (in the ABCD matrix formalism), which means that the CAM can be reduced to an effective two mirror cavity consisting of the planar end mirror and a curved micro-mirror array.  \textbf{b.} In this effective cavity picture, the mode size at the micro-mirror array (and MLA) increases monotonically as $\Delta d_{Mirror}$ is swept from zero to $f_{\textrm{MLA}}$. \textbf{c.} Asphere stability diagrams at three values of $\Delta d_{\textrm{Mirror}}$, illustrating that  $\Delta d_{\textrm{Mirror}}$ fixes the maximum size of the cavity mode waist (as the asphere separation is varied). At zero asphere defocus (dashed line), the mode waist is exactly equal to $w_{\textrm{MLA}}/M$.}
	\label{fig:waist_heuristics}
\end{figure}

In Eq.~\eqref{eq:1} of the previous section, we showed that the mode waist of the CAM is set by $f_\textrm{MLA}$ and $M$, as long as $\Delta d_{\textrm{Mirror}}=f_{\textrm{MLA}}/2$. In this section we derive Eq.~\eqref{eq:1} and generalize our understanding to different values of $\Delta d_{\textrm{Mirror}}$ by systematically reducing the CAM's physics to that of a simple two mirror cavity. Finally, we explain how $\Delta d_{\textrm{Mirror}}$ can be tuned in conjunction with the asphere position to select a maximum mode waist for the CAM.

To derive Eq.~\eqref{eq:1}, we begin with the constraints enforced by the CAM's 4f telescopes; the primary demagnifying telescope sets a linear relationship for both the mode sizes and mode spacings between the plane of the MLA and central image plane: $w_0=w_{\textrm{MLA}}/M$, where $w_{\textrm{MLA}}$ is the mode size at the MLA, $w_0$ is the mode waist, and $M$ is the telescope magnification. 
Importantly, 4f telescopes are imaging systems which return rays exactly onto themselves after a double-pass: the action of a round trip through the two 4f telescopes is equal to an identity matrix in the ABCD matrix formulation~\cite{siegman1986lasers}. As such, we can create a model of the CAM that removes the 4f telescopes and replaces them with a ``virtual'' end mirror right behind the MLA without changing the (paraxial) behavior of the light rays. The combination of virtual end mirror and MLA acts identically to a micro-mirror array with mirror radii of curvature $R=f_\textrm{MLA}$, allowing us to effectively reduce the CAM to an array of half-planar two-mirror cavities each consisting of the planar in-coupling mirror and an $R=f_{\textrm{MLA}}$ curved mirror (Fig.~\ref{fig:waist_heuristics}a). 

The stability diagram of half-planar cavities is simple\textemdash depending only on the mirror spacing and radius of curvature~\cite{siegman1986lasers}. For a spacing of $\Delta d_{\textrm{Mirror}}=f_{\textrm{MLA}}/2$, the mode waist on the planar mirror is $(f_{\textrm{MLA}}\times\lambda/(2\pi))^{1/2}$ and the mode waist on the curved mirror is $w_{\textrm{MLA}}=(f_{\textrm{MLA}}\times\lambda/\pi)^{1/2}$. Combining this expression with the telescope demagnification condition yields Eq.~\eqref{eq:1}.

Now we consider what happens when $\Delta d_{\textrm{Mirror}}$ is adjusted. In the picture of the effective half-planar cavity, when $\Delta d_{\textrm{Mirror}}$ is close to zero, the mode waist on both mirrors is infinitesimally small. As  $\Delta d_{\textrm{Mirror}}$ is increased, the mode size on the curved mirror grows monotonically, approaching infinity at $L=f_{\textrm{MLA}}$ (Fig.~\ref{fig:waist_heuristics}b). Because the 4f telescope simply demagnifies the MLA plane, the CAM mode waist in the image plane obeys the same trend with $w_0=w_\textrm{MLA}/M$.

Up to this point, we have maintained the telescope in a perfect 4f configuration while changing $\Delta d_{\textrm{Mirror}}$; however it is crucial to understand how displacements of the aspheric lens factor in, since the asphere stability diagram is sensitive over $\upmu$m length scales.
In Fig.~\ref{fig:waist_heuristics}c, we choose three different values of $\Delta d_{\textrm{Mirror}}$ and plot the CAM waist with respect to displacements of the aspheric lens. These aspheric lens displacements slightly defocus the 4f telescope. At zero defocus, the waist is equal to $w_0=w_\textrm{MLA}/M$, as expected. However, for $\Delta d_{\textrm{Mirror}}>f_{\textrm{MLA}}/2$ ($\Delta d_{\textrm{Mirror}}<f_{\textrm{MLA}}/2$) the maximum waist is shifted to a point where the telescope has a slight positive (negative) defocus. It is thus possible to choose $\Delta d_{\textrm{Mirror}}$ to achieve larger (or smaller) cavity mode waists, and then longitudinally displace the aspheric lens to a position where this new waist is maximally stable.

It is important to be able to infer this maximum waist when aligning the CAM. Within the paraxial approximation, the width of the asphere stability diagrams (for longitudinal displacements) is equal to twice the Rayleigh range of the maximum waist. In the lab, this provides a valuable diagnostic tool for quickly inferring the mode waist and corresponding value of $\Delta d_{\textrm{Mirror}}$, and tuning them to the desired value (Sup.~\ref{SI:Alignment}).

\section{Cavity Field of View}
\label{sec:fov}
\begin{figure*}[t] 
    \includegraphics{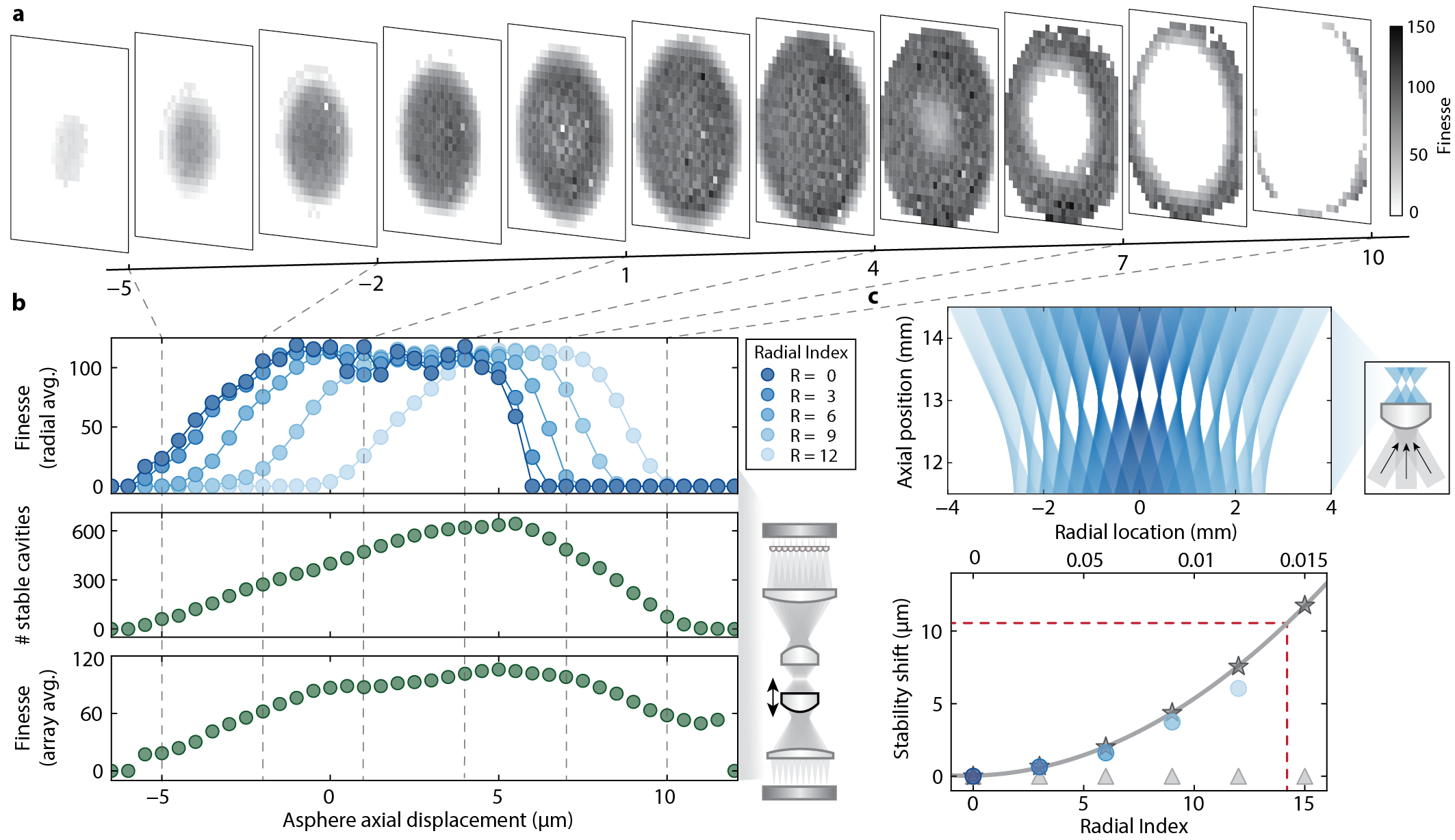}
    \caption{
        \textbf{Field of View.} \textbf{a, b.} Both (\textbf{a}) the field of view\textemdash plotted as an array of cavity finesses\textemdash and (\textbf{b, top}) the stability of cavities at a given radial displacement from the central cavity are sensitive to micron-scale longitudinal displacements of the two aspheres. Radially averaged finesse measurements are used as a proxy for assessing the corresponding stability regions, with lines connecting points to guide to the eye. We work at a displacement that simultaneously maximizes both number of stable cavities (\textbf{b, middle}) and the array-averaged finesse (\textbf{b, bottom}); with further optics development, we anticipate substantial gains in the achievable field of view. \textbf{c.} The field curvature of the asphere (\textbf{top}) produces a radially dependent, quadratic focal shift which translates the asphere stability regions as a function of radial cavity index, limiting the field of view. (\textbf{inset}) We visualize this curvature by ray tracing tilted plane waves through a single pass of the aspheric lens. (\textbf{bottom}) Experimental (circles), and numerically calculated (stars, see Sup.~\ref{SI:RayTracing})) stability diagram shifts align with the expected focal shift from the asphere field curvature (multiplied 4 for the number of asphere passes per cavity round-trip), while numerical calculations using ideal lenses (triangles) exhibit no shift. We attribute the small discrepancy between the experimental data and the numerics to quadratic misalignment aberrations, such as defocus, which add to or subtract from the intrinsic field curvature. Red lines indicate the radial index and stability shift corresponding to the maximum field of view. 
        }
	\label{fig:fov}
\end{figure*}
One of the primary figures of merit for the CAM is its intrinsic field of view (FOV), which we define as the diameter in the central image plane over which simultaneously stable, near-diffraction-limited cavity modes are formed. The number of achievable cavities is related to the field of view as $N_{cavities}\approx\pi(d_\textrm{FOV}/2d_0)^2$ where $d_0$ is the spacing of the mode waists in the central image plane. 

In this realization, the intrinsic FOV ($d_\textrm{FOV}$) is limited to $280~\upmu$m (corresponding to $\approx 615$ cavities) by the field curvature of the aspheric lens, discussed below, while the total observable set of cavities is physically limited to a diameter of $300~\upmu$m due to a ring piezo on the incoupling mirror. We expect that the intrinsic field of view can be greatly expanded by using wide field of view, high NA microscope objectives in place of the aspheric lenses~\cite{manetsch2024tweezer}.

In the previous section, we characterized the stability of the central cavity with respect to perturbations of the CAM's most sensitive optic\textemdash the aspheric lens. Here we use this method to experimentally and numerically extract stability regions for all cavities in the CAM. In Fig.~\ref{fig:fov}a, we plot the CAM field of view for a set of aspheric lens longitudinal displacements over a range of 18~$\upmu$m. Strikingly, we observe that the central cavities become stable first, followed by those at larger radial indices. After a displacement of 11~$\upmu$m\textemdash equal to the expected width of the stability region\textemdash the central cavities become unstable, leaving a ring-like structure that gets thinner as the innermost cavities become unstable. Radially averaged slices of this data explicitly shows these shifted stability regions (Fig.~\ref{fig:fov}b, top).  To determine the ideal operating point of the CAM we plot the number of stable cavities and array averaged finesse (Fig.~\ref{fig:fov}b, middle and bottom), finding that the array-averaged finesse and number of cavities are maximized at similar aspheric lens displacements, corresponding to the point right before the central and outermost stability regions become non-overlapping. 

In order to quantitatively understand the source of the radially-dependent stability shifts, we performed a set of numerical ray-traced simulations of the CAM (Fig.~\ref{fig:fov}c, bottom). For a CAM composed of ideal lenses, no stability shift is observed (triangles); in contrast, simulations incorporating realistic aspheres (stars) exhibit a stability shift consistent with the experimentally measured stability shifts (circles), with residual deviations attributable to additional first order aberrations, such as defocus, which arise from alignment imperfections rather than from intrinsic properties of the system. Other aberrations, including astigmatism and spherical aberration are found in our numerics to have a negligible impact (Sup.~\ref{SI:Astigmatism}, Sup.~\ref{SI:SphericalAberr}).

The stability shifts uncovered by these numerics are attributable to the field curvature of the aspheric lenses, 
illustrated in the top panel of Fig.~\ref{fig:fov}c. The numerically calculated stability shifts quantitatively follow the single-pass ray-traced field curvature of the asphere, multiplied by the number of passes through the aspheric lens per round trip (Fig.~\ref{fig:fov}c, bottom, solid gray line). Within the CAM, multiple round trips through the optics constrain all cavity waists to lie in a flat central image plane, and thus the asphere field curvature instead manifests as shifts to individual cavity stability regions which ultimately limit the CAM field of view.

In order to expand the CAM field of view and avoid the deleterious effects of field curvature, the aspheres could be replaced with high NA, wide field of view microscope objectives. Such objectives would have a few unique design constraints: (1) the single-pass loss should be kept below $\sim 1\%$ to ensure high cavity array finesse; and (2) beyond maintaining a flat focal plane, the objective must ensure equal absolute path lengths to all field points, as is required for the degeneracy condition (Sec.~\ref{sec:degeneracy}). High NA objectives in typical atom array experiments are already designed to provide diffraction-limited performance with sub-micron field curvature across their specified field of view. With minimal additional design, we anticipate that a 1.5~mm field of view objective at 10~$\upmu$m spacing could realize over 17,000 simultaneously resonant cavities, with uniform mode waists, provided that the loss condition can be met~\cite{manetsch2024tweezer}.

\section{Cavity Finesse}
\label{sec:finesse}
Finesse is one of the primary figures of merit of the CAM, determining, for a fixed waist, the cavity cooperativity and thus the light collection efficiency~\cite{tanji2011interaction}. In this section we take a detailed look at the observed and expected cavity losses, and their cavity-to-cavity variation.

\begin{table}[t]
    \centering
    \caption{\textbf{Cavity Loss Budget.} Single-pass loss from each individual optic is listed, alongside the number of passes for one cavity round-trip. The ``measured'' loss of each intra-cavity optic was determined by placing it between two highly reflective end mirrors and measuring the resulting finesse at low NA. The loss from the end mirrors was provided by the manufacturer. The ``expected'' loss was calculated from the manufacturer-provided AR coating reflectivity and from scattering losses. Scattering loss was calculated via the normal-incidence total integrated scatter in transmission~\cite{bennett1961relation}: TIS $\approx (\frac{2\pi\sigma}{\lambda}(n-1))^2$ with $\lambda=780$~ nm, where $\sigma$ is the RMS surface irregularity given by the manufacturer. The surface roughness from all spherical optics is sub-nanometer and thus negligible. Total element-wise losses are calculated as $1-\prod_i (1-l_i)^p$, where $l_i$ is the component loss, and $p$ is the number of passes.}
    \vspace{2mm}
    \begin{tabular}{l l l c}
        \toprule
        \textbf{Optic} & \textbf{\makecell[l]{Measured\\Loss (\%)}} & \textbf{\makecell[l]{Expected\\Loss (\%) \ \ \ }} & \textbf{\makecell{Passes\\}} \\
        \midrule
        Incoupling mirror   & 0.1 & 0.1 & 1\\
        Microlens array     & 0.1 & 0 & 2\\
        Spherical lens 1    & 0.07 & 0 & 2\\
        Aspheric lens 1     & 0.8 & 0.42 & 2 \\
        Aspheric lens 2     & 0.8 & 0.42 & 2\\
        Spherical lens 2    & 0.07 & 0 & 2\\
        Outcoupling mirror  & 0.1 & 0.1 & 1\\
        \midrule
        \textbf{Total (element-wise)\ \ }    & \textbf{3.8} & \textbf{1.9}\\
        \textbf{Total (from peak finesse)}   & \textbf{4.2} & \\
        \textbf{\makecell[l]{Total (from avg. finesse)}} & \textbf{5.3} & \\
        \bottomrule
    \end{tabular}
    \label{etab:losses2}
\end{table}

Due to the CAM's constituent intra-cavity optics, there are a multitude of potential sources of loss that can degrade the finesse, including scattering from surface roughness, specular reflection and absorption from AR coatings, clipping on lens apertures, and dust. Here we catalog losses from each optic, finding that anti-reflective (AR) coating and scattering losses dominate, and then detail strategies for mitigating these sources and others. 

In Table~\ref{etab:losses2}, we summarize the measured and predicted losses for each intra-cavity optic, as well as across the array. 
The experimentally measured average round-trip loss per cavity is 5.3\%, corresponding to the array averaged finesse of 114. However, there are individual cavities with poor performance, which we attribute to dust, local imperfections on the MLA and mirrors, and radially dependent clipping loss (most prominent for cavities operated near the edge of their stability region). In order to disambiguate the intrinsic losses (due to AR coatings and surface roughness) from the extrinsic losses (due to dust and clipping), we identify the cavity with the highest finesse (145), and use its corresponding loss (4.2\%) as a baseline for comparison against expected coating and scattering losses.

We compute the intrinsic loss using two methods: first, from a combination of the manufacturer-provided AR coating curves and calculated scattering loss due to surface roughness; second, from a direct measurement of the loss of each intra-cavity optic by placing it inside of a simple two-mirror ``characterization'' cavity of known finesse. We combine these losses to obtain an element-wise total measured loss of 3.8$\%$ (Table~\ref{etab:losses2}). We find that the element-wise loss is consistent with the measured peak finesse, differing by a small amount (0.4$\%$) which could be accounted for by minimal dust accumulation on the intra-cavity lenses or imperfect alignment of the full composite system.

The predominant source of loss in the CAM comes from the intra-cavity aspheric lenses, which each have 0.8\% measured single-pass loss and thus account for more than three quarters of the total intra-cavity loss. About half of this loss can be explained from the manufacturer-provided AR coating reflectivity and surface roughness (Table~\ref{etab:losses2}). We suspect that the remaining $\sim0.4\%$ loss comes from non-idealities in the AR coating, as the aspheric lenses are highly curved which makes it difficult to both deposit a coating with uniform thickness and ensure stable performance across all angles of incidence.

In addition to the aspheric lenses, a small amount of loss is added by the other intra-cavity optics\textemdash the two spherical lenses and the MLA. These spherical optics have negligible scattering loss due to their exceptionally low surface roughness. Each of these optics are coated with a high quality anti-reflective V-coating centered at 780~nm. In theory these coatings have zero reflectance at their design wavelength, however in practice we measure 0.07\% single-pass loss for the spherical lenses and 0.1\% loss for the MLA. 

Degradation from dust is also a major concern. For cavity optics in the multiple image planes of the CAM, such as the end mirrors and MLA, dust and imperfections result in isolated cavities with low finesse. For cavity optics in and near Fourier planes, including the aspheric and spherical lenses, dust impacts all cavities to a similar degree because the cavity modes at the lenses are large and overlapping. To minimize loss from dust, we find that a positive pressure, HEPA filtered enclosure is sufficient to avoid noticeable degradation over months-long time periods. Without HEPA filtered air, we observe the array-averaged loss of the CAM increase by several percent in just a couple weeks. Fortunately, the optics can be cleaned and restored to their initial performance with standard cleaning techniques~\cite{phelps2010first}.

A final source of loss in the CAM is clipping on the apertures of the intra-cavity optics. In the design of the CAM, we take precautions to minimize total clipping losses to $<0.1\%$, a small amount compared to the total $5.3\%$ array-averaged loss. The primary source of clipping loss is the high NA asphere. Keeping the total clipping loss to a sub-$0.1\%$ level requires choosing a mode waist such that the $1/e^2$ mode diameter at the lens is no larger than half of the lens diameter. For our NA$=0.55$ asphere, a mode waist around 1~$\upmu$m is an optimum below which clipping losses begin to significantly degrade the finesse. Poor alignment of the CAM can also contribute to additional clipping loss and can be mitigated through careful alignment according to the procedure detailed in Sup.~\ref{SI:Alignment}.

There are several ways in which careful selection of the intra-cavity optics could further reduce the amount of intrinsic loss. For the aspheric lens, a high index glass could be used to reduce its curvature, making it possible to coat the asphere with a more performant AR coating\textemdash though such glasses often have high absorption. For applications that can tolerate larger mode waists, the aspheric lenses could even be replaced by spherical lenses which can have losses below $0.1\%$. For applications requiring small mode waists and large fields of view, we believe that an intra-cavity high NA microscope objective consisting of V-coated or nanotextured optics could achieve sub-$1\%$ loss, similar to that of our asphere. Finally, losses due to the MLA could be eliminated entirely through a slight modification of the CAM geometry: replacing the incoupling mirror and microlens array with a micro-mirror array~\cite{ding2025highfinesse}. With these further improvements in optic design and AR coatings, we expect that total losses could be reduced to $\approx1.5\%$, corresponding to array averaged finesses over 400, and a $3.5\times$ improvement in cooperativity.

\section{Cavity degeneracy}
\label{sec:degeneracy}

Besides the number of cavities and their finesses, the third figure of merit for the cavity array microscope (CAM) is the array degeneracy, or the degree to which all cavities can be made simultaneously resonant. The cavity frequency is controlled by the round-trip optical path length of each cavity's local axis and Gouy phase. In particular, the shift in resonance frequency between a cavity with indices $(i,j)$ and the central one depends on both the change in one-way optical path length $ \Delta L(i,j)$, as well as on the difference in round-trip Gouy phase $\Delta \phi_{rt}(i, j)$: 
\begin{equation}
    \Delta \nu = \nu_{FSR} \left( \frac{1}{\pi} \Delta \phi_{rt} - \frac{2 \Delta L}{\lambda}\right)
    \label{eq:gouy}
\end{equation}
Experimentally, this is measured as the fractional detuning, $\Delta_m(i,j) = \Delta \nu/\ \nu_{\textrm{FSR}}$.

One of the principal benefits of the CAM architecture is that simultaneous resonance can be achieved purely by proper positioning of the intra-cavity optics, without explicit requirements for local cavity control. For a CAM with ideal optical components, this condition is met when all lenses are perfectly centered transversely along the same optical axis, and when the aspheric and spherical lenses are positioned longitudinally in a perfect telescope configuration. This is the case because, in the paraxial approximation, a 4f imaging system not only images points to points, but also plane-waves to plane-waves, thereby ensuring that all path lengths through the system are equal.

\begin{figure}[t!] 
    \includegraphics{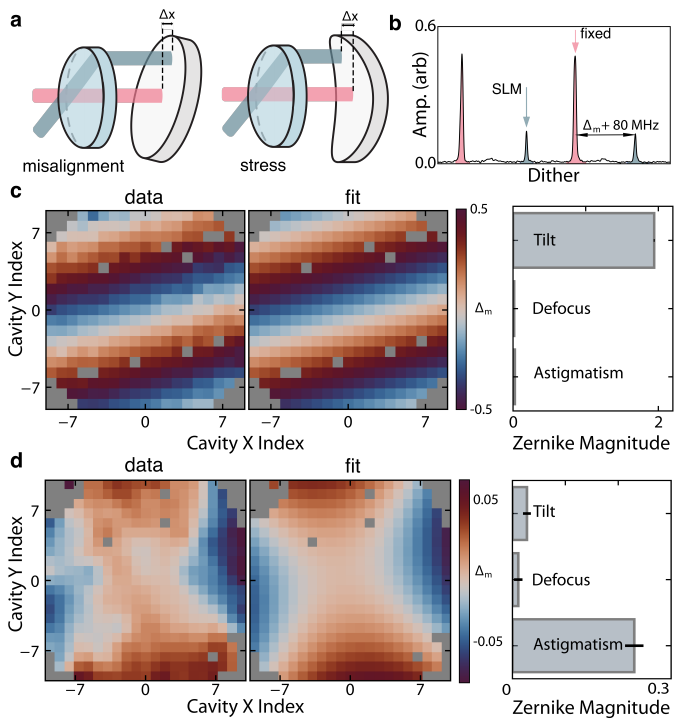}
    \caption{\textbf{Degeneracy characterization.} \textbf{a.} The degeneracy of the CAM is sensitive to perturbations that change the relative round-trip optical path length of the individual cavities in the array. Such perturbations can, for instance, arise from misalignments of various intra-cavity optics, or from imperfections of their surfaces, such as stress or form-error of the end mirrors. \textbf{b.} To characterize these effects, we measure the fractional detuning $\Delta_m$ of each cavity (blue) relative to the central cavity (red). \textbf{c.} Transverse misalignment of the spherical lens produces a linear gradient of detunings across the array, while \textbf{d.} a mechanically stressed end-mirror produces astigmatism across the array. We distinguish these effects by decomposing the map of detunings into a sum of Zernike polynomials (with one site Gaussian smoothing applied on the data), allowing quantitative evaluation of aberrations, order-by-order. Error bars denote the bootstrapped standard error of the fitted Zernike coefficients.}
	\label{fig:degen_align}
\end{figure}

In practice, various non-idealities break this condition. Misalignments of the various cavity optics can change the relative optical path lengths in predictable ways. For instance if one of the end mirrors is tilted, or if one of the spherical lenses is transversely offset from the optical axis, then a linear detuning gradient will be applied across the array. Similarly, aberrations in optics arising from 
surface irregularity or mounting-stress can lead to differences in path length (Fig.~\ref{fig:degen_align}a). Finally, the Gouy phase variation across the array is a substantial fraction of $\pi$ resulting in an additional radially dependent frequency shift (Sup.~\ref{SI:gouy}). 

We interrogate the cavity degeneracy with a fixed excitation beam coupled into the central cavity and a variable excitation beam controlled with an SLM 
(Fig.~\ref{fig:exp_setup} and Fig.~\ref{fig:degen_align}b). The fixed excitation beam is detuned from the SLM beam by 80~MHz (approximately half the FSR) and the relative powers are tuned such that the two drives are unambiguously distinguishable on simultaneous photodiode measurements of the cavity transmission (Fig.~\ref{fig:degen_align}b). From this photodiode signal we can extract the relative cavity detuning, and by addressing different cavities across the array with the SLM we can measure the detuning across the entire array modulo the cavity FSR. Intentionally misaligning an optic (e.g. tilting an end mirror) or using a known-astigmatic end mirror produces detuning maps reflecting the expected aberration (Fig.~\ref{fig:degen_align}c, d). Crucially, we are able to decompose these detuning maps into a sum of low-order Zernike polynomials, thereby quantitatively identifying the degree to which different aberrations affect the array degeneracy (Fig.~\ref{fig:degen_align}c, d, right).

Ultimately, by carefully controlling optic alignments and aberrations we are able to realize hundreds of simultaneously degenerate cavities, detuned by less than the optimal outcoupling linewidth (Fig.~\ref{fig:summary}d). In the following, we describe in detail the techniques and heuristics we employ to perform this optimization, the sensitivity to misalignments, and the means by which we can use the CAM to identify aberrations in its own optics at the nanometer-scale.

\subsection{Misalignments of the cavity imaging system}


\begin{figure}[t!] 
    \includegraphics[width=\columnwidth]{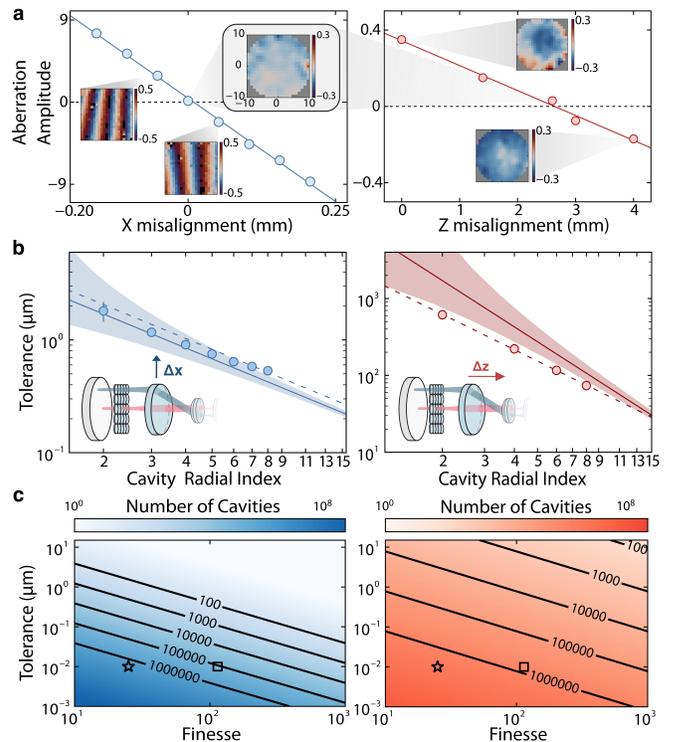}
    \caption{\textbf{Positioning of the intra-cavity optics.} \textbf{a.} Achieving a degenerate CAM requires precise alignment of intra-cavity optics. (\textbf{left}) Transverse or (\textbf{right}) longitudinal misalignments of the $\textrm{f}=300$~mm spherical lens produce linear (i.e. ``tilt'') or quadratic (i.e. ``defocus'') variations, respectively, in the round-trip optical path length across the array, observed as variations in site frequencies. We fit each detuning map using Zernike polynomials, plotting the dominant Zernike coefficient as a function of lens displacement and operating where the coefficient crosses zero for all axes. For visual clarity, the inset detuning maps are displayed with one site Gaussian smoothing. \textbf{b.} Positional tolerance (microns of spherical lens transverse and longitudinal displacement required to shift the relative cavity detuning by one linewidth) as a function of cavity $x$-index ($i$, with $j$ = 0). The solid line indicates the theoretical model developed in Eq.~\ref{eq:3} with the shaded region indicating the range of deviation that would occur if the chosen center cavity differed from the central optical axis by $\pm1$ (dark red/blue) or $\pm2$ (light red/blue) cavity sites. Error bars on individual data points denote propagated uncertainties from the fitted slope of detuning versus lens displacement. \textbf{c.} The number of cavities that can be made simultaneously resonant depend on the tolerance with which the spherical lens can be positioned and the cavity finesse. At our array averaged finesse of 114, and with the $\approx 10$~nm precision of our translation stage (Thorlabs MAX381), this corresponds to over $>10^5$ cavities (square). For the readout-optimized linewidth (Sup.~\ref{SI:collection}), this corresponds to $>10^6$ cavities (star).
    }
	\label{fig:degen_sensitivity}
\end{figure}

Achieving degeneracy across hundreds of cavities requires precise positioning of the optics in the CAM's 4f telescopes. As illustrated in Fig.~\ref{fig:degen_sensitivity}a, transverse and longitudinal misalignments of the CAM's spherical lens induce linear (tilt) and quadratic (defocus) non-degeneracies. Other degrees of freedom, such as end mirror tip/tilt, affect the degeneracy in a similar manner\textemdash in fact we observe that all misalignment-induced non-degeneracy is predominantly linear and quadratic. This means that, after careful pre-alignment of all optics (Sup.~\ref{SI:Alignment}), small residual non-degeneracies can be compensated entirely with the spherical lens position, regardless of the culpable optic. 

This compensation is also important for correcting the fractional detuning induced by differential Gouy phases across the array. The Gouy phase varies radially through a significant fraction of $\pi$ due to the fact that we increase the size of our array by maximizing the set of cavities with intersecting stability regions. The resulting spread of Gouy phases is approximately quadratic with radial index and can therefore be corrected via mm-scale longitudinal adjustments of the spherical lenses (Sup.~\ref{SI:gouy}). We note, however, that the possible negative effects of an over-reliance on this compensation are not fully explored. In the following, we hone in on the misalignment-induced non-degeneracies\textemdash deriving their analytical form, assessing their positional tolerance, and elucidating the surprising impact of the MLA on the degeneracy landscape.

Misalignment-induced non-degeneracies arise due to small optical path length differences across the CAM. In order to better understand these, we derive a paraxial model of fractional detuning $\Delta_m$ (assuming zero differential Gouy phase), as a function of both transverse ($\mathbf{\delta_r}=\delta_x\mathbf{\hat{x}}+\delta_y\mathbf{\hat{y}}$), and longitudinal ($\delta_z$) displacements of the demagnifying, $\textrm{f}=300$~mm, spherical lens from its ideal telescope position. We compute $\Delta_m$ for a generalized off-axis cavity indexed by $(i,j)$, located at a transverse position $\mathbf{r}=p_\textrm{MLA}(i\mathbf{\hat{x}}+j\mathbf{\hat{y}})$ in the image plane of the spherical lens where $p_\textrm{MLA}$ is the MLA pitch. This yields

\begin{equation}\Delta_m(\mathbf{\delta_r},\delta_z,\mathbf{r}) = \frac{1}{\lambda f}\left(\frac{\delta_z||\mathbf{r}||^2}{f}-2~ \mathbf{\delta_r}\cdot \mathbf{r}\right)
    \label{eq:2}
\end{equation}
for a spherical lens of focal length $f$. Equation~\ref{eq:2} is consistent with our observation that longitudinal misalignments $(\delta_z)$ of the spherical lens result in quadratic non-degeneracies across the array, while transverse misalignments produce linear non-degeneracies (Fig.~\ref{fig:degen_sensitivity}a). Moreover, the dependence on $\mathbf{r}$ indicates that cavities located further from the central optical axis are increasingly sensitive to misalignment. These results define a set of alignment tolerances specifying the maximum allowable lens displacement ($\delta_x,~\delta_y,~\delta_z)$ before a given cavity becomes detuned from the central cavity by one linewidth. We obtain closed form expressions for the spherical lens positional tolerance per fractional linewidth by separately considering each term in Eq.~\ref{eq:2}, solving for $\delta_{x,y,z}$ and normalizing by the array-averaged finesse, $F=114$:

\begin{equation}(\delta_x,~\delta_y,~\delta_z)=\frac{\lambda f}{F}\left(\frac{1}{2p_\textrm{MLA}\times i}~,~\frac{1}{2p_\textrm{MLA}\times j}~,~\frac{f}{||\mathbf{r}||^2}\right)
\label{eq:3}
\end{equation}

In Fig.~\ref{fig:degen_sensitivity}b, we experimentally measure this tolerance as a function of cavity x-index ($i$, with $j=0$) for both transverse and longitudinal displacements
of the spherical lens. The dashed line indicates the theoretical tolerance predicted by Eq.~\ref{eq:3}, while the shaded region denotes the associated uncertainty arising from a potential offset between the chosen central cavity and optical axis of up to $\pm1$ cavity site. Additionally, the low sensitivity of the degeneracy to $z$ displacements means that the coupling between axes can result in a lower measured tolerance than theoretically predicted (Sup.~\ref{SI:tolerance}).

These tolerance plots provide insights into the conditions required to operate at degeneracy, showing that in order to stabilize this 600 cavity CAM to degeneracy, the spherical lens transverse position must be controlled to within 300~nm, while the longitudinal position only needs to be controlled to within 10~$\upmu$m. Such levels of precision are readily achievable using closed loop feedback with precision 3-axis piezo stages (e.g. Thorlabs MAX381, which we employ here). 

We also investigate the theoretical limit for the number of cavities that can be simultaneously stabilized at degeneracy when limited only by the positional accuracy of our translation stage. As illustrated in Fig.~\ref{fig:degen_sensitivity}c, this fundamental limit is over $10^5$ cavities (and $>10^6$ cavities at the ideal outcoupling linewidth), implying that other factors\textemdash most notably the CAM field of view\textemdash will constrain the achievable number of cavities count well before degeneracy stabilization becomes a limiting factor.

\begin{figure}[t!] 
    \includegraphics{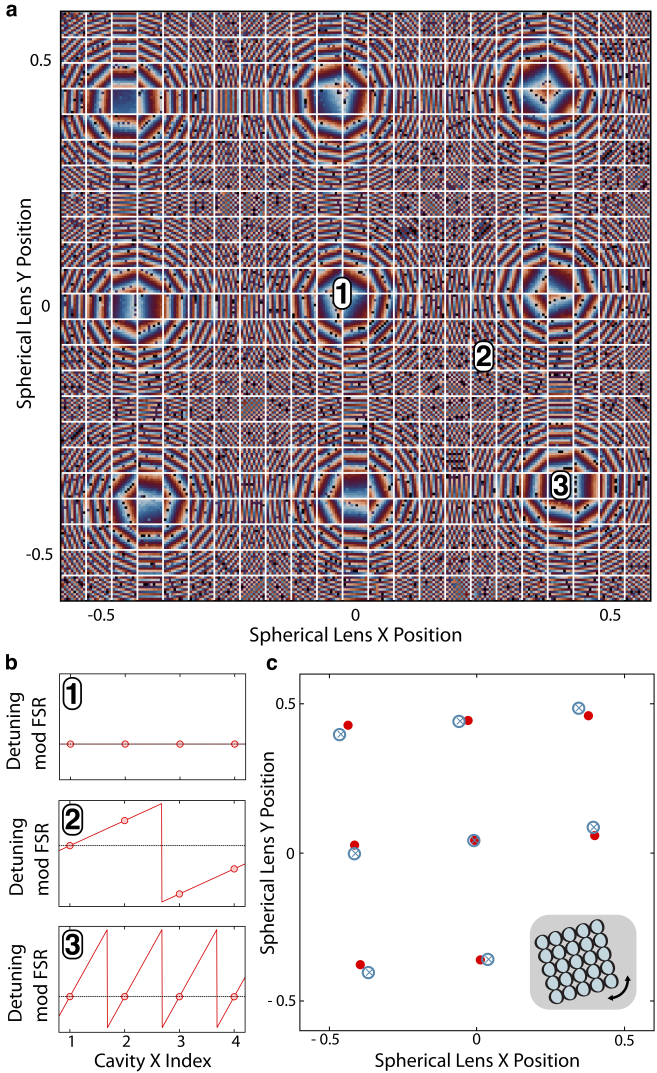}
    \caption{
        \textbf{Aliasing of the degenerate condition.} \textbf{a.} We measure detuning maps for multiple choices of spherical lens displacement along both transverse axes ($x, y$), revealing a varied landscape: at many lens positions, the array appears degenerate, leading to ambiguity as to what the correct alignment is. \textbf{b.} We interpret these effects as arising from aliasing due to the finite spacing of the MLA which samples the underlying displacement-induced linear phase gradient, depicted with cartoon plots of detuning mod FSR. \textbf{c.} To identify the non-aliased alignment, we rotate the MLA and plot the locations of the fitted degeneracy points from \textbf{a,} both with (blue) and without (red) a four degree rotation; for the rotated data, we compare the measured degeneracy points (blue crosses) and a rotated prediction (blue open circles) based off the original points in red, finding good agreement. We identify the "true" degeneracy point as the center of rotation about which other "fake" degeneracy points rotate.
        }
	\label{fig:degen_alias}
\end{figure}

Having determined the required tolerance for the spherical lens positioning, we now identify and characterize the degeneracy condition in further detail. Naively, Eq.~\ref{eq:2} suggests that there is a unique degeneracy point where $\Delta_m(0,0,\mathbf{r}) = 0$ for all cavity positions $\mathbf{r}$. However, due to the periodicity of each cavity's longitudinal mode spectrum, we can only distinguish fractional detunings $\Delta_m(\mathbf{\delta_r},\delta_z,\mathbf{r})\mod 1$. This leads to an array of seemingly identical degeneracy points at different displacements of the spherical lens, with the spacing determined by the MLA pitch, lens focal length, and probe wavelength. We experimentally observe this extended grid of degenerate configurations in Fig.~\ref{fig:degen_alias}a, where each sub-panel shows the detuning map for a given transverse displacement ($\delta_x,~\delta_y)$ of the spherical lens (with the longitudinal displacement fixed near $\delta_z=0$).

To understand the extended grid of degenerate points, we revisit the treatment of the cavities' longitudinal modes. For a set of cavities with identical optical path lengths, the longitudinal spectra across the cavities will be degenerate. However, if the one-way optical path length varies linearly with an integer multiple of $\frac{\lambda}{2}$ between each cavity, they will also be degenerate. Thus, the extended grid of degenerate points reflects conditions wherein the magnitude of the linear detuning error induced by the transverse displacement of the spherical lens causes multiples of $\frac{\lambda}{2}$ optical path length variation between adjacent cavities in a grid established by the regular spacing of the MLA (Fig.~\ref{fig:degen_alias}b). 

To set a reproducible alignment target and mitigate the effect of potential off-axis aberrations, we seek to park the spherical lens at the ``true'' degenerate point where $\Delta_{\nu}(0,0,\mathbf{d}) = 0$ \textit{and} the central axes of all optics are co-aligned. We identify this true degeneracy point by rotating the MLA, thereby changing the spherical lens transverse positions at which the periodic condition above is met. After fitting the positions of the degenerate points for both the rotated and unrotated case, we see that all of the degenerate points rotate according to the MLA angle about a center point coinciding with the true degeneracy location Fig.~\ref{fig:degen_alias}c. We park the spherical lens at this location, leaving further investigations of the negative impacts associated with using the other degeneracy points for future studies.

\subsection{Resolving nanometer-scale aberrations}

\begin{figure}[t!] 
    \includegraphics[width=\columnwidth]{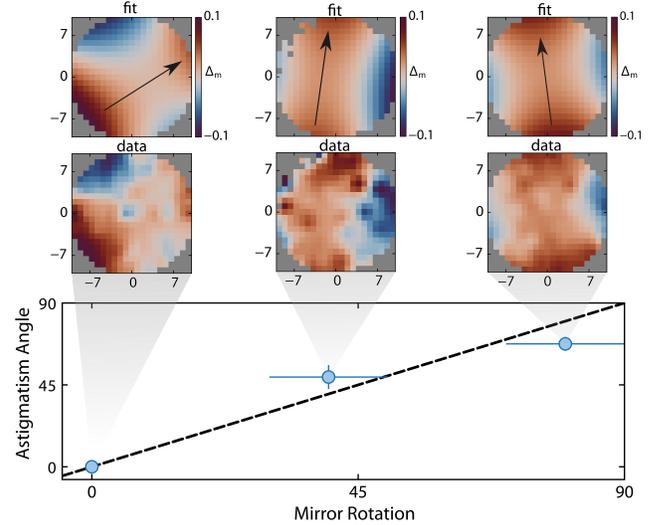}
    \caption{\textbf{Aberration characterization.} Measuring cavity degeneracy allows us to evaluate the intrinsic aberrations of specific optics. For instance, we insert a known astigmatic mirror as the end-mirror of the cavity, and then measure the angle of the fitted astigmatism as the mirror is rotated in place, observing a close correlation. Horizontal error bars denote the uncertainty in the measured mirror rotation angle; vertical error bars denote the standard error propagated from the bootstrapped Zernike astigmatism coefficient.}
	\label{fig:degen_astig}
\end{figure}

In addition to the bulk adjustments necessary to approach the degeneracy point, residual non-degeneracies arise from imperfections, including stress on optics and surface irregularities which introduce variations in the optical path length between cavities. The CAM is most sensitive to displacements and surface errors of optics located in/near image planes of the circulating light, i.e. where the independent cavity modes are spatially separated, as this causes differential\textemdash rather than common-mode\textemdash changes in path length.

We study this for the case of a known-aberrated planar mirror for which glue-induced mounting stress leads to tens of nanometers of curvature of the mirror surface. Measuring the map of detunings as described above allows the CAM to directly characterize this surface curvature as residual non-degeneracy once the effects of the bulk misalignments from Fig.~\ref{fig:degen_sensitivity} have been removed by tuning to degeneracy. After swapping the unstressed planar end mirror for the stressed sample, we measure the detuning map at three rotational orientations of the stressed mirror, and extract the resulting rotation of the astigmatism from the fitted Zernike polynomials (as in Fig.~\ref{fig:degen_align}d), finding good agreement between the two (Fig.~\ref{fig:degen_astig}).

We can apply this procedure to analyze general variations in surface profile and aberrations, so long as the aberrated optics are located in/near an image plane of the cavity-array (so as to have an impact on the optical path length) and as long as the magnitude of the induced error is larger than the limits on resolution. At an average finesse of $F = 114$, the linewidth of the resonator sets a path length resolution of $\frac{\lambda}{2 F} = 3.4$~nm (smaller than the observed $\approx40$ nm peak-to-peak variation of the aberrated mirror) with further limits imposed by residual non-degeneracies from other optics.

\begin{figure}[t!] 
    \centering
    \includegraphics[width=81mm]{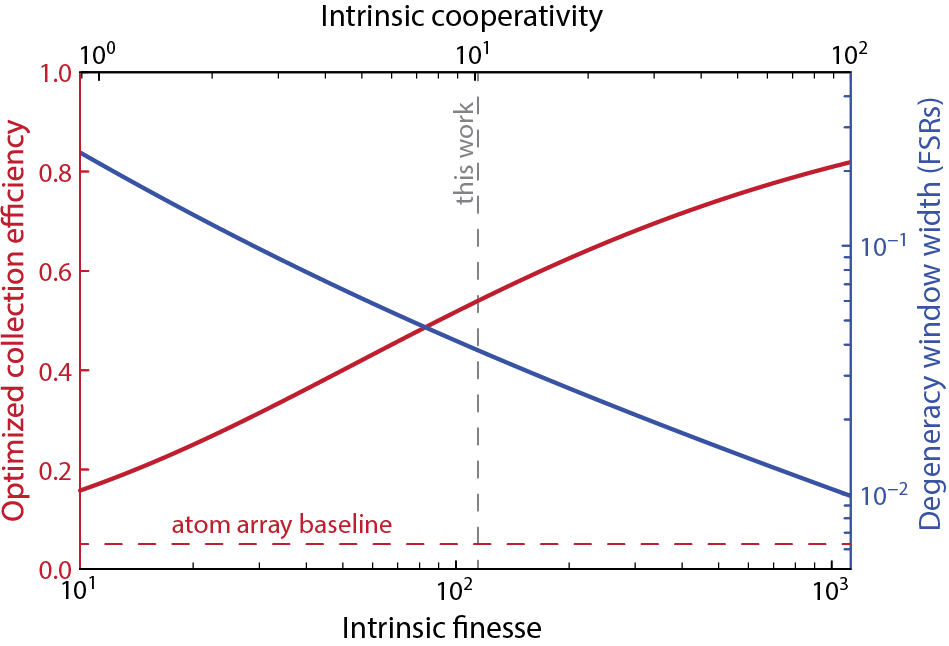}
    \caption{\textbf{Projected cavity performance.} Total optimized light collection probability (solid red) for current CAM parameters, see Sup.~\ref{SI:collection}; our projected performance is already an order-of-magnitude greater than the $\approx 5\%$ expected for standard atom arrays (dashed red), with further room to develop as the intrinsic cavity finesse is improved. Notably, even as the finesse grows, the acceptable window of cavity non-degeneracy (i.e. red bars in the color bar of Fig.~\ref{fig:summary}d) decreases only mildly (solid blue).}

	\label{fig:collection}
\end{figure}

\section{Outlook}
\label{sec:outlook}

The cavity array microscope (CAM) is a powerful new technology for controlling light-matter interaction across an array of optical cavities. In this paper we have identified the fundamental mechanisms governing its scalability and demonstrated the largest, most performant CAM to date, with roughly 600 cavities operating with intrinsic finesse greater than 100 and mutual degeneracy within the outcoupling optimized linewidth. Combined with a similarly sized neutral atom quantum processor node, we expect that this architecture will enable record quantum networking bandwidth, while minimally disturbing computing operations. A critical component of this work is our quantitative analysis of imperfections in the CAM. Through insights gained in these studies, we anticipate a straightforward path towards tens of thousands of mutually degenerate cavities with intrinsic finesses approaching one thousand, and correspondingly improved networking performance~\cite{sinclair2025faulttolerant} (Sup.~\ref{SI:degenlock},~\ref{SI:finesse_opt}).

Besides networking, the CAM will also dramatically improve readout rates of neutral atom quantum processors, approaching $\sim 10~\upmu$s readout with our projected $53\%$ collection efficiency (Fig.~\ref{fig:collection}, Sup.~\ref{SI:collection}). By inducing diffractive coupling between cavities, we expect to realize coherent coupling of photons between neighboring cavity sites, enabling studies of hitherto unexplored light-matter Hamiltonians~\cite{makin2008quantum,hartmann2008quantum,greentree2006quantum}. Further enhanced connectivity could be enabled by introducing programmable optics (like deformable mirrors) into the system, enabling fast, parallelized, non-local atom-atom interaction within a single array. If instead of collecting photons from the atoms, the CAM is instead used to transduce environmental photons \textit{onto} the atom array, it could realize a natural constituent of a quantum-computing-enhanced optical imaging system~\cite{mokeev2025enhancing}.

We emphasize that while we have designed this current iteration of the CAM around compatibility with neutral atom quantum processors, the fundamental apparatus is versatile, and may enable gains across a variety of disciplines, e.g. biological or chemical sensing~\cite{reynolds2025quantum,needham2024labelfree}. Ultimately, the CAM is the culmination of a long, active, and fruitful history of cavity-engineering; through studying, characterizing, and optimizing the factors which affect its performance (as we have done here), we anticipate a similarly fruitful new frontier of many-cavity QED awaits.

\section{Acknowledgements}
This work was supported by AFOSR grant FA9550-22-1-0279, ARO grant W911NF-23-1-0053, and AFOSR MURI Grant FA9550-19-1-0399. D.S. acknowledges support from the NSF GRFP. A.S. acknowledges support from the Hertz Foundation, as well as the NDSEG Fellowship under AFOSR grant FA9550-23-F-0014. A.L.S. acknowledges support from the Stanford Science Fellowship, and additionally by the Felix Bloch Fellowship and the Urbanek-Chodorow Fellowship. We thank Brandon Grinkemeyer for fruitful discussions regarding networking bandwidth. 

\section{Competing Interests}
D.S., D.I.S. and J.S. hold a patent on the resonator geometry demonstrated in this work. J.S. is a consultant for, and on the advisory board of, Atom Computing.

\bibliographystyle{naturemag}
\bibliography{references}

\clearpage


\subsection{Data Availability}
The experimental data presented in this manuscript are available from the corresponding author upon request, due to the proprietary file formats employed in the data collection process.
\subsection{Code Availability}
The source code for simulations throughout are available from the corresponding author upon request. 
\subsection{Additional Information}
Correspondence and requests for materials should be addressed to J.S. (jonsimon@stanford.edu). Supplementary information is available for this paper.

\clearpage
\newpage

\onecolumngrid
\newpage
\section*{Supplementary Information}
\appendix
\renewcommand{\appendixname}{Supplement}
\renewcommand{\theequation}{S\arabic{equation}}
\renewcommand{\thefigure}{S\arabic{figure}}
\renewcommand{\figurename}{Supplemental Information Fig.}
\renewcommand{\tablename}{Supplemental Information Table}
\setcounter{figure}{0}
\setcounter{table}{0}
\numberwithin{equation}{section}

\section{Ray tracing numerics}
\label{SI:RayTracing}

While paraxial ABCD matrix calculations are a valuable tool for understanding the stability of the central cavity of the CAM, all other off-axis cavities are inherently non-paraxial, with no way to compute their stability regions in the fully paraxial paradigm. In order to interrogate the stability of all cavities we move beyond the paraxial approximation and use ray tracing to generate local ABCD matrices that provide stability information across the array.

For the numerical simulations in Fig.~\ref{fig:fov}d of the main text, as well as Sup.~\ref{SI:SphericalAberr}, Sup.~\ref{SI:Astigmatism}, and Sup.~\ref{SI:gouy}, we use a home-built 4D ($x$ position, $x$ slope, $y$ position, $y$ slope) numerical ray tracer, which allows us to interrogate the stability diagrams across all cavities in the CAM. For Fig.~\ref{fig:fov}d and Sup.~\ref{SI:Astigmatism} we simplify the ray-tracing by approximating the two spherical lenses and the constituent lenses of the MLA as idealized paraxial optics under the ABCD formalism. For the asphere, however, such an ABCD formalism breaks down due to non-paraxial effects associated with the high NA, thus requiring Snell's law to trace rays through the custom aspheric lens profile. Conversely, for Sup.~\ref{SI:SphericalAberr}, we treat the aspheres paraxially in order to isolate the specific effect of the spherical lenses. 

Our process for computing the cavity array stability diagrams is as follows: first, we identify the central axis of the cavity of interest via gradient descent. Next we propagate two test-rays\textemdash with small angular and radial displacements relative to the cavity axis\textemdash through one round trip of the CAM, starting in the central plane between the two lenses. We determine the elements of the round-trip ABCD matrix from the displacement between the output test-rays and cavity axis. Finally, we use this effective round-trip ABCD matrix to determine the cavity's effective eigen-$q$ parameter from which we extract the cavity mode waist.

\section{Spherical aberration}
\label{SI:SphericalAberr}
\begin{figure*}[t!] 
    \includegraphics[width=172mm]{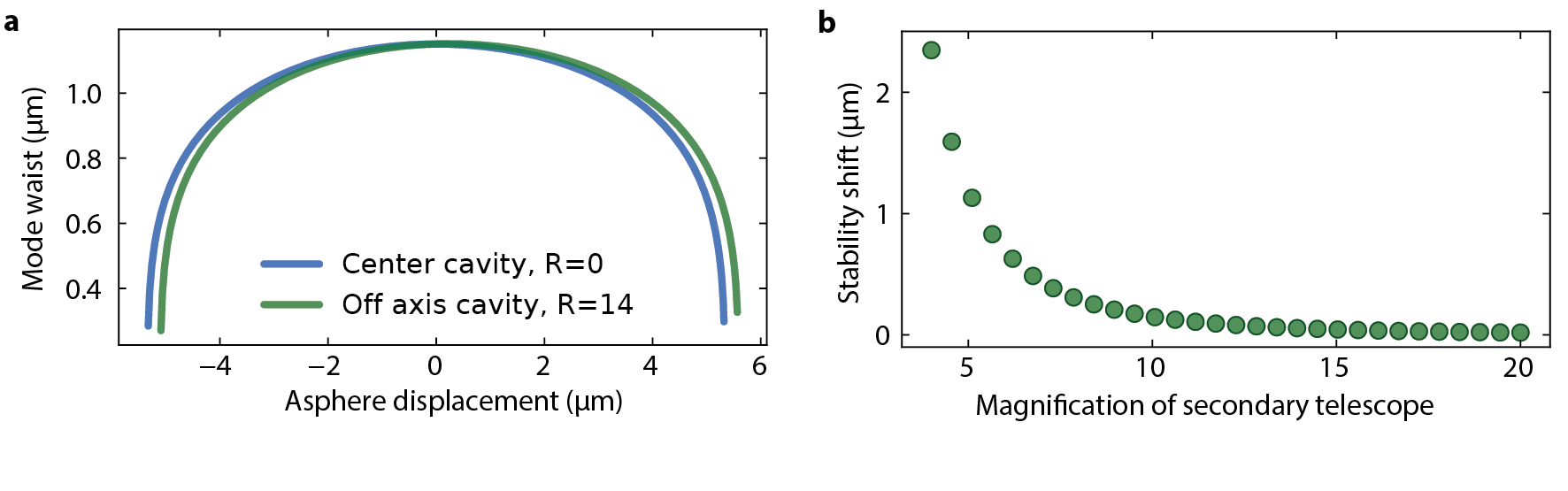}
    \caption{\textbf{Spherical aberration. a.} Spherical aberration shifts the stability region of a cavity at a radial index of 14 (green) by 0.25~$\upmu$m relative to the central cavity (blue). \textbf{b.} As the focal length of the $\textrm{f}=200$~mm spherical lens is reduced, the magnitude of spherical aberration-induced stability shift increases substantially.}
	\label{fig:spherical_aberration}
\end{figure*}

In order to construct the telescopes of the CAM with the desired magnification factor, we need large EFL spherical lenses. These lenses are desirable due to their extremely low surface roughness and correspondingly low loss. Although we operate these lenses at low NA, spherical aberrations can still come into play because the off-axis cavities necessarily pass through the lenses at positions displaced from the lens' optical axis. To quantify the resulting effect, we perform numerical ray tracing simulations in which all intra-cavity components are treated as ideal except for the two spherical lenses. We find that the effects of spherical aberration are small and non-limiting for our current system. The maximum spherical aberration-induced stability shift\textemdash occurring for cavities at the edge of the field of view\textemdash is only 0.25~$\upmu$m (Fig.~\ref{fig:spherical_aberration}a). This shift, however, becomes more problematic for shorter spherical lens focal lengths (for fixed CAM waist) due to the associated increase in NA.

To illustrate this effect, we perform additional simulations in which we vary the magnification of the secondary telescope via the focal length of the $\textrm{f}=200$~mm spherical lens. We hold the $\textrm{f}=300$~mm spherical lens and $\textrm{f}=10$~mm asphere focal lengths constant in order to maintain a fixed waist. As the focal length is reduced, the stability shift increases rapidly, underscoring the importance for short focal length lenses to be corrected for operation at higher NA.

\section{Astigmatism}
\label{SI:Astigmatism}
\begin{figure*}[t!] 
    \includegraphics[width=172mm]{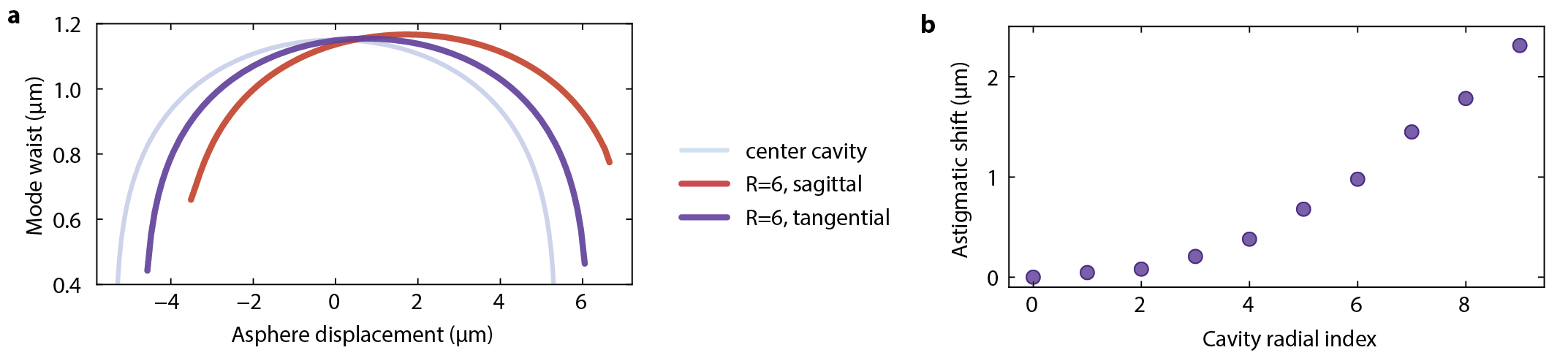}
    \caption{\textbf{Astigmatism. a.} The CAM exhibits substantial astigmatism due to the aspheric lenses. At a radial index of 6, the sagittal (pink) and tangential (purple) stability regions are split out by over a micron, in addition to the overall shift relative to the central cavity (light blue) due to field curvature. \textbf{b.} The degree of astigmatism (as measured via the relative shift between the sagittal and tangential stability regions) increases with radial index.}
	\label{fig:astigmatism}
\end{figure*}

In the main text, we explained how the field curvature of the aspheres is the primary aberration limiting the field of view of the CAM. Astigmatism is a related aberration that causes differing amounts of field curvature between the sagittal and tangential foci of a Gaussian beam passing through a lens at an angle relative to the central axis. The sagittal focus occurs from rays propagating in the plane formed by the central axis of the lens and central axis of the incoming Gaussian beam, while the tangential focus from rays propagating in an orthogonal plane.

To illustrate the effect of this astigmatism on the CAM stability, we numerically ray tracing the CAM in 3D using the exact surface profiles of the aspheric lenses, and plot the resulting sagittal and tangential stability diagrams. As expected, for the center cavity, we observe no shift between sagittal and tangential stability diagrams, whereas for a cavity displaced from the center by 6 sites, we observe 1~$\upmu$m of splitting (Fig.~\ref{fig:astigmatism}a). Because the shift is never larger than the width of the stability diagram within our field of view, it is not a limiting effect. To resolve both the field curvature and astigmatism, high NA, wide field of view objectives should be used, as they are typically designed to limit both field curvature and astigmatism within their quoted field of view.

\section{Gouy phase across the array}
\label{SI:gouy}

\begin{figure*}[t!] 
    \includegraphics[width=172mm]{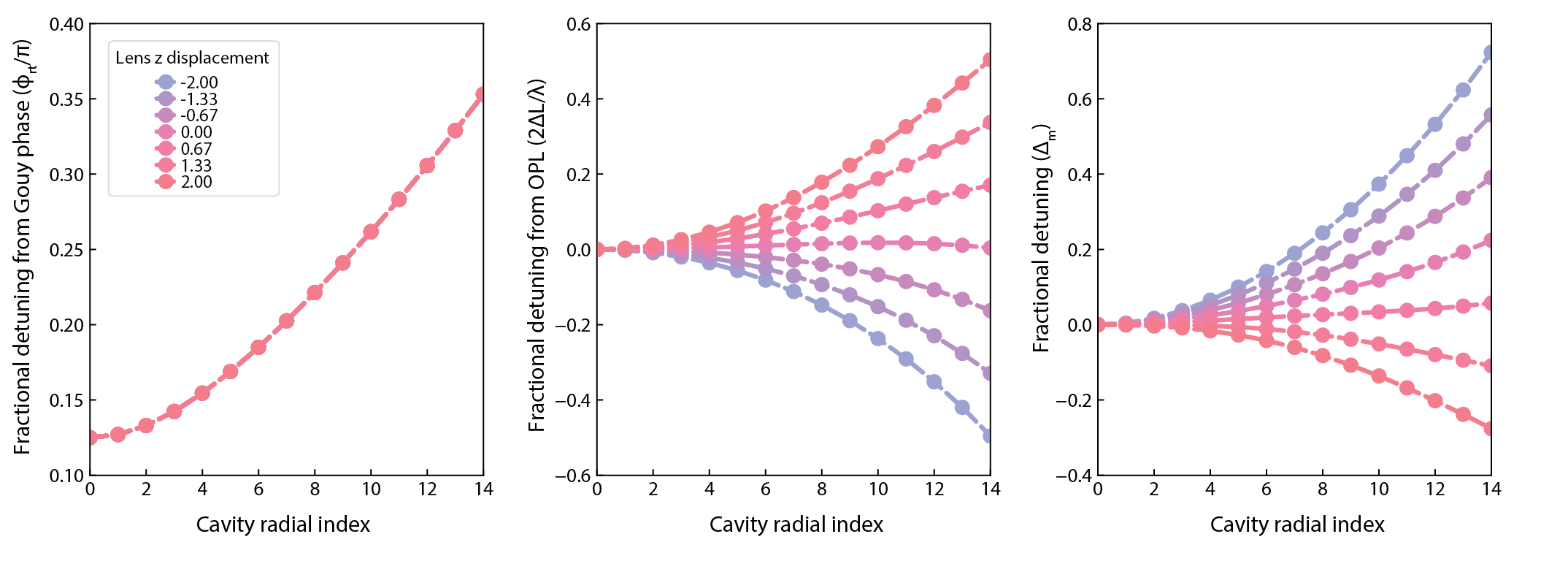}
    \caption{\textbf{Gouy phase and optical path length.} The fractional detuning of a cavity from the center ($\Delta_m$) depends on both the Gouy phase, $\phi_{rt}$, and optical path length (OPL) difference, $\Delta L$. \textbf{a.} The Gouy phase of each cavity varies quadratically with radial index, leading to a detuning across the array that spans a significant fraction of the cavity FSR. The spread of Gouy phases is the same for all displacements of the spherical lens, after slighly adjusting the asphere to set center cavity to a Gouy phase of $\pi/8$. \textbf{b.} Displacements of the spherical lens create a quadratic variation in OPL as a function of cavity radial index. In a perfect telecentric configuration the OPL is independent of cavity index, but in the real system, spherical and higher order aberrations lead to deviations at large radii. \textbf{c.} Because both the Gouy phase and the OPL are quadratic in the cavity radius, the spherical lens $z$ position can adjusted by $\approx 0.6$~mm in order to compensate the two effects and achieve degeneracy over a large number of cavities.}
	\label{fig:gouy}
\end{figure*}

Gouy phase is a generic property of Gaussian beams, accrued when the beam passes through a focus and measured as the phase relative to that of a perfect plane wave. Because the field of view (FOV) of our CAM is set entirely by radially-dependent shifts of the cavity stability region induced by non-paraxial beam propagation through the aspheric lens (Sec.~\ref{sec:fov}), the Gouy phase varies across the array (at fixed alignment) over a range bounded between 0 and $\pi$. 

This spread of Gouy phases across the CAM factors into the CAM degeneracy, adding up t o 1 FSR of additional detuning between the center and outermost cavities (see Eq.~\ref{eq:gouy} and Fig~\ref{fig:gouy}a). The Gouy phase variation is largely quadratic in radial index, and thus correctable with a small amount of telescope defocus. To demonstrate this, in Fig.~\ref{fig:gouy}b, we plot the relative detuning due to the optical path length differences accrued when the telescope is slightly defocused (achieved by displacement of the $\textrm{f}=200$~mm spherical lens). At small radial indices the detuning varies quadratically, with higher order deviations at large radial indices due to spherical and higher order aberrations. At a defocus of $\approx0.6$~mm, we can largely cancel out the detuning induced by the Gouy phase shift across the array, allowing us to achieve degeneracy over a large number of cavities (Fig.~\ref{fig:gouy}c). 

In order to maximize the FOV at each spherical lens displacement, we always set the Gouy phase of the center cavity to $\approx\pi/8$ via $\upmu$m-scale displacements of the asphere.

\section{Optimal collection efficiency}
\label{SI:collection}

The CAM can dramatically improve light collection efficiency and related tasks (e.g. networking rates), but optimal light collection is not found at the same condition as maximum finesse and cooperativity. Instead, assuming small overall loss, the light collection efficiency is nominally given by 
\begin{align}
\eta=\frac{C}{1+C}\frac{L_o}{L_o+L_i}\,,
\end{align}
where $C$ is the cooperativity, $L_o$ is the probability of photon transmission through the outcoupling mirror and $L_i$ is the round-trip loss for the photon excluding the outcoupling mirror. Here we assume that the photons are generated by weak driving or vSTIRAP so that spontaneous emission from the excited state is negligible. We can recast all variables in terms of the finesse, as $F\approx2\pi/(L_o+L_i)$ and $C=\alpha\times F$ (where $\alpha=(6/\pi^3)(\lambda/w)^2$ with wavelength $\lambda$ and focused beam waist $w$~\cite{tanji2011interaction}). If we further define the intrinsic ($L_o=0$) finesse as $F_i=2\pi/L_i$ with associated intrinsic cooperativity $C_i$, we can solve for $\eta$ as a function of $F_i$ and $L_o$, and by further optimizing over $L_o$ we find the following compact expression for the maximum light collection efficiency for a given intrinsic finesse, assuming optimal outcoupling,
\begin{align}
\eta^*=\frac{\sqrt{1+C_i}-1}{\sqrt{1+C_i}+1}\,,
\end{align}
with optimal outcoupling mirror transmission of
\begin{align}
L_o^*=\frac{2\pi\sqrt{1+C_i}}{F_i}\,,
\end{align}
and optimized finesse
\begin{align}
F^*=\frac{F_i}{1+\sqrt{1+C_i}}\approx\sqrt{\frac{F_i}{\alpha}}-\frac{1}{\alpha}\,.
\end{align}

These scaling laws allow us to predict the performance of both our current and future CAM realizations (Fig.~\ref{fig:collection}). Notably, for our current parameters, $\alpha\approx0.09$ and $F_i=114$, we predict an optimized collection efficiency of $53\%$ at an outcoupling mirror transmission of $18\%$, with a resultant optimized finesse of $F^*\approx26$, corresponding to a cavity linewidth of 5.5~MHz and cooperativity of 2.3. This optimized finesse sets the deviation window of cavity detunings which can be read out (given by $1/F^*$), and notably increases only as the square root of the intrinsic finesse, meaning much higher intrinsic finesses and corresponding collection efficiencies can be achieved with only a modest restriction on acceptable spread of cavity detunings.

\section{Birefringence}
\label{SI:Birefringence}
\begin{figure}[t!] 
    \includegraphics[width=90mm]{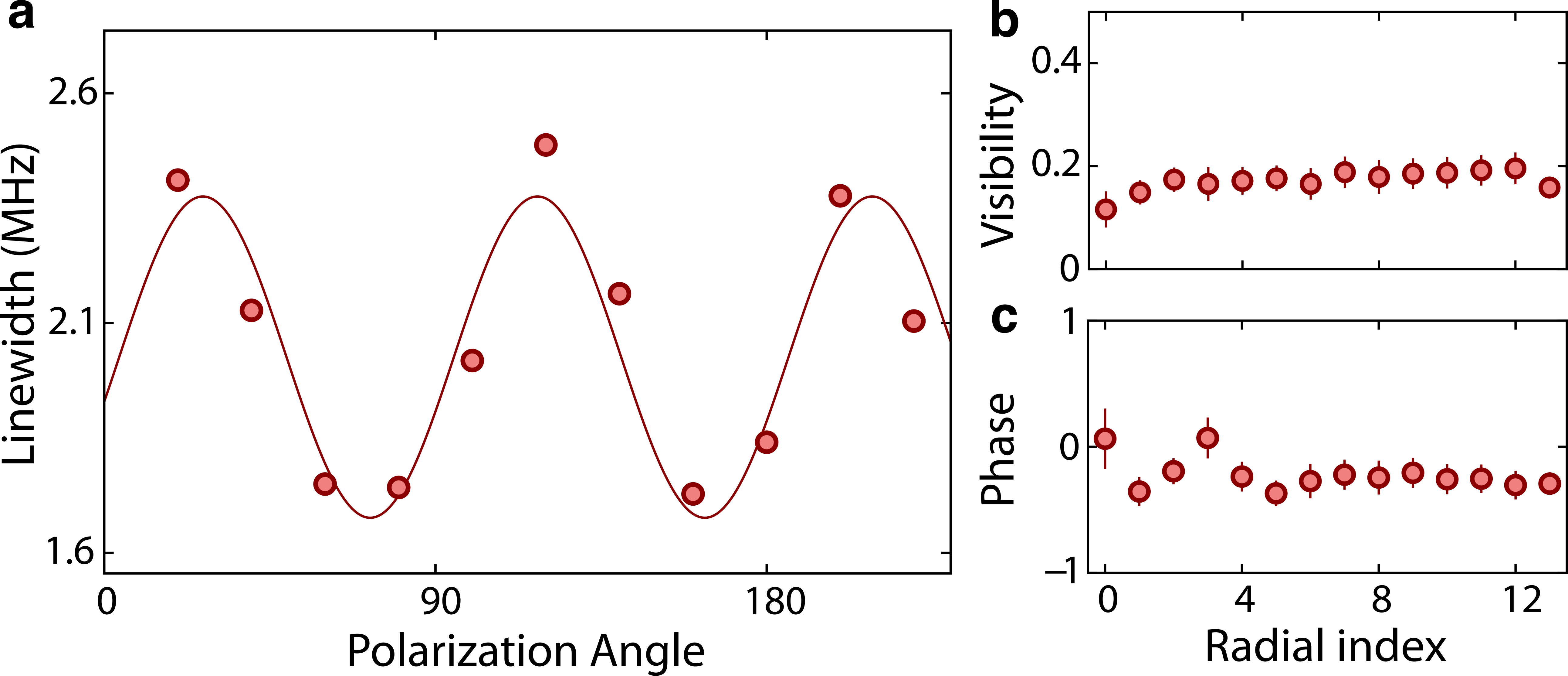}
    \caption{\textbf{Cavity birefringence}. \textbf{a.} Effective cavity linewidth for the central cavity as the input linear polarization angle is varied. We observe a birefringent variation, but no resolvable splitting. We believe the observed birefringence is primarily caused by the dielectric coatings on the aspheres, due to the difficulty of engineering a non-birefringent coating for the highly curved (maximum AOI of $\approx 50$ degrees) surfaces. \textbf{b, c.} We do not see a significant variation in the visibility, defined as the amplitude of the fitted sine function divided by its offset, or in the phase (radians) of the oscillations relative to that of the center cavity. }
	\label{fig:birefringence}
\end{figure}

Birefringence is an important consideration for all cavity systems, as symmetry breaking between polarization modes can degrade performance of imaging or networking protocols that rely on polarization encoding. The CAM is potentially sensitive to such effects due the AR coatings of its intra-cavity, optics. To test the birefringence, we vary the input (linear) polarization angle and observe the resulting cavity spectra. 

We do not observe a resolvable splitting of cavity polarization modes, but do see that the cavity linewidth oscillates with the polarization angle (Fig.~\ref{fig:birefringence}a), indicating a potential mode splitting at the level of $<0.005$ FSRs. Notably, this behavior is consistent across the whole array, and we do not see substantial variation in either phase or visibility of the oscillations as a function of cavity radial index (Fig.~\ref{fig:birefringence}b, c). Because the splitting is far less than the optimal outcoupling linewidth (Sup.~\ref{SI:collection}), we anticipate that it will be minimally intrusive for preliminary networking demonstrations, and expect that it can be further reduced by using nanotextured coatings instead of dielectric AR coating. Additionally, a fixed birefringence could be compensated by an intra-cavity phase-retarder engineered to entirely cancel the effect, a benefit of the free-space nature of the CAM.

\section{Measuring tolerance}
\label{SI:tolerance}
\begin{figure}[t!] 
    \includegraphics[width=160mm]{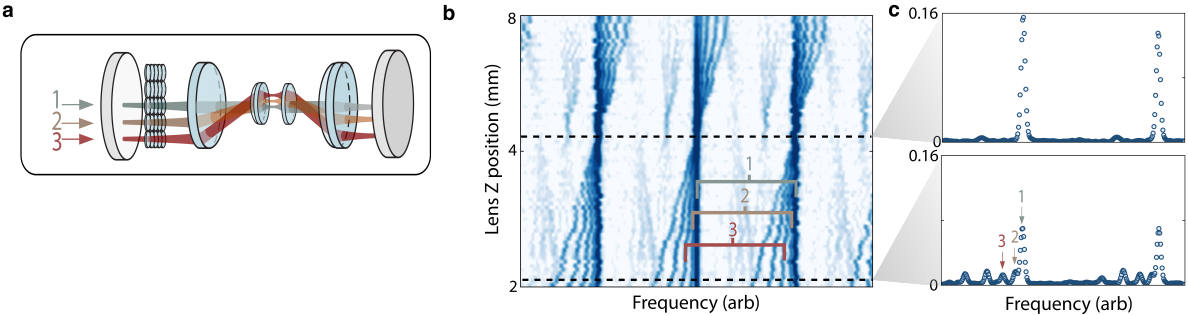}
    \caption{\textbf{Cavity resonances}. \textbf{a.} We drive multiple cavities in a row simultaneously to measure the longitudinal degeneracy tolerance of the $f=300$~mm spherical lens  Specifically, cavity "1" has a radial index of 0 (center cavity), cavity "2" has a radial index of 2, and cavity "3" has a radial index of 4, etc. \textbf{b.} Detunings of five cavity spectra as the longitudinal ($z$) position of the lens is scanned. The cavities begin detuned from each other and come into resonance as the position of the spherical lens is brought towards the optimal alignment position. \textbf{c.} Zoom outs of individual cavity spectra at (\textbf{top}) the optimal lens $z$ position where the frequencies of the five cavities are degenerate, and (\textbf{bottom}) at a misaligned $z$ position where the five cavities are separated in frequency.}
    \label{fig:waterfall}
\end{figure}

Measuring the positional tolerance of intra-cavity optics is crucial for assessing the degeneracy condition and prospects for active stabilization. 
Due to experimental constraints, we use slightly different procedures to measure the tolerance to transverse ($x$) and longitudinal ($z$) misalignments. For transverse misalignments, we extract tolerances from the detuning maps introduced in Fig.~\ref{fig:degen_align}, tracking the relative detunings of a row of cavities while the spherical lens is translated in the $x$ direction. Although the same procedure can in principle be applied to longitudinal misalignments, the limited tuning of the degeneracy to longitudinal misalignments over the available travel range of the lens's translation stage makes this data particularly sensitive to shaking and residual transverse misalignments.

To reduce susceptibility to shaking, we employ a different technique with higher data rate. We again track the change in detunings across a row of cavities but now drive five cavities (of radial indices of 0, 2, 4, 6, and 8) simultaneously (Fig.~\ref{fig:waterfall}a). In order to disambiguate the five resonances, the center cavity is set to a high power, while the four off-axis cavities are set to roughly equal lower power. Ultimately, this allows the relative detunings to be rapidly extracted at hundreds of spherical lens $z$ positions. In Fig.~\ref{fig:waterfall}b, we display this data as a map of raw cavity traces as the spherical lens $z$ position is swept through degeneracy. The traces in Fig.~\ref{fig:waterfall}c provide snapshots of the constituent traces both at and away from degeneracy. The additional faint features in between main peaks correspond to the higher order modes from each of the five cavities, which also come into resonance at the optimal degeneracy point. 

We extract the tolerances (shown in Fig~\ref{fig:degen_sensitivity}b) from this data by using prominence-based peak search to track the position of each resonance as the $z$ position of the spherical lens is misaligned. Although this method provides rapid data acquisition, its validity is limited due to cross talk between the $z$ and $x/y$ axes, which makes the tolerance appear lower than its theoretical value. 

\section{Performance over time}
\label{SI:stability_time}
\begin{figure*}[t!] 
    \includegraphics[width=100mm]{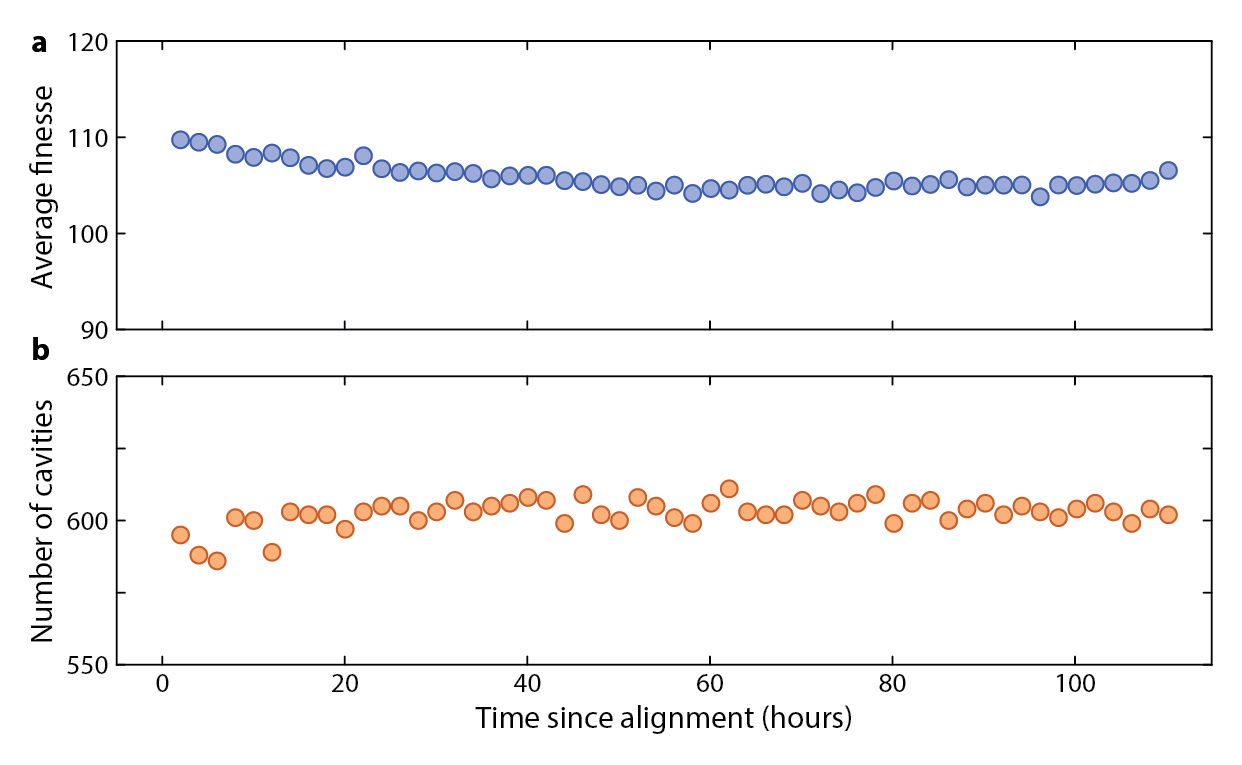}
    \caption{\textbf{Cavity drifts}. The cavity retains its performance for many days following alignment, both in terms of \textbf{a.} the mean finesse, as well as \textbf{b.} the total number of resolvable cavities.}
	\label{fig:stability_v_time}
\end{figure*}
Stability of the number of cavities and finesse across the CAM is critical for long-term operation. In Fig.~\ref{fig:stability_v_time}a, b, we monitor the array-averaged finesse and number of cavities over the course of four days without any adjustment to the constituent optics. 

Variations in the number of cavities can be caused by drifts in the axial separation of the two aspheric lenses, as illustrated in Fig.~\ref{fig:fov}. The small variation ($\approx \pm 15$) in cavity number indicates that the separation drifts by less than $\pm1~\upmu$m. This is sufficient for maintaining continuous operation of 95$\%$ of the cavities in the CAM and even stabler operation is expected when the lenses are placed in vacuum. However, a tighter tolerance could be achieved by using a slow digital feedback loop that monitors the transverse mode splitting of a single cavity as a proxy for the axial separation of the aspheric lenses.

The CAM array-averaged finesse exhibits negligible variation over that timescale as well. As highlighted in Section~\ref{sec:finesse}, a HEPA filtered enclosure is critical for long term operation in order to avoid dust accumulation on intra-cavity optics.

\section{Locking to degeneracy}
\label{SI:degenlock}

\begin{figure*}[t!] 
    \includegraphics[width=180mm]{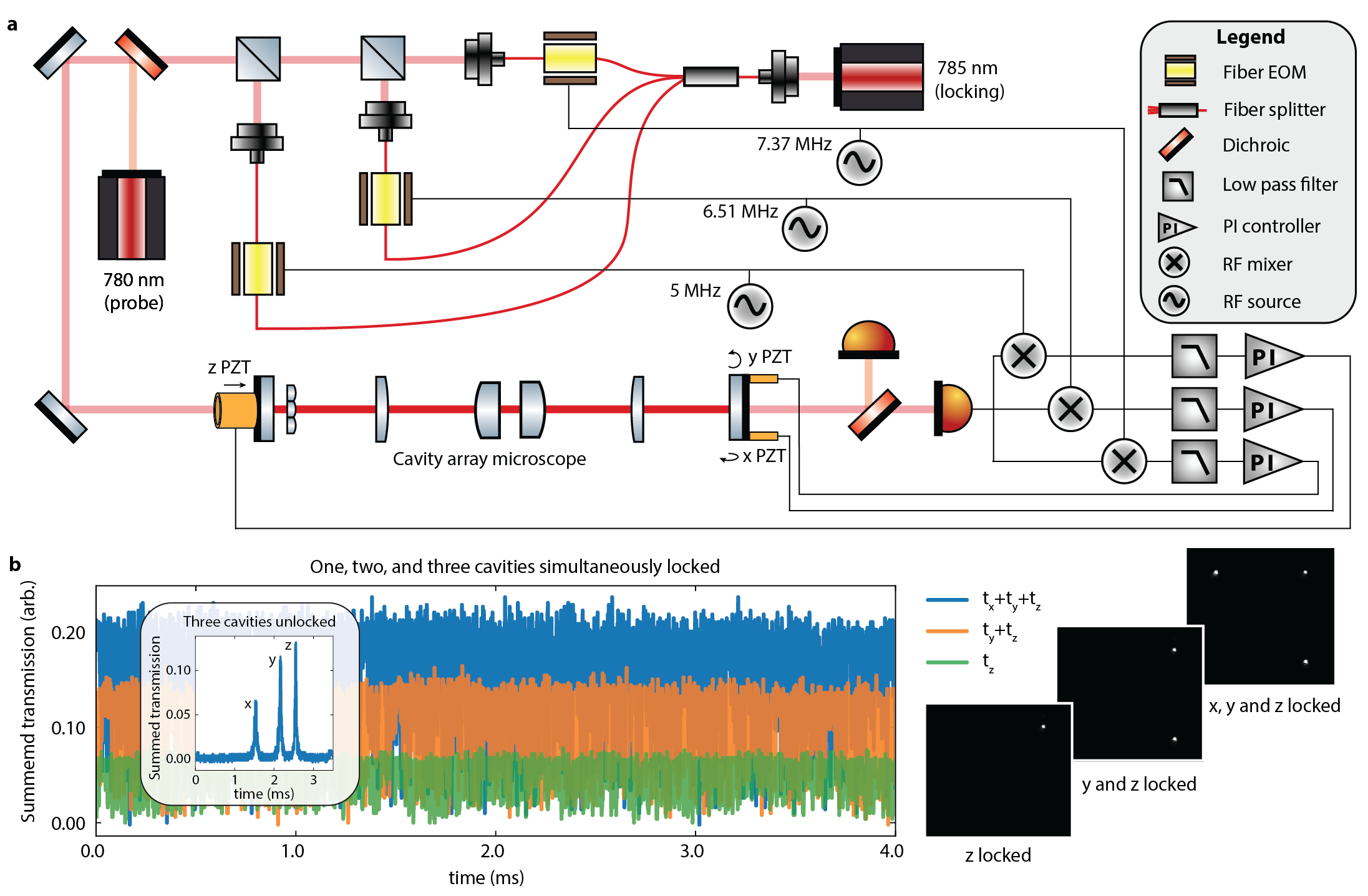}
    \caption{\textbf{3-axis locking}. Simultaneous resonance of all 600 cavities is maintained using PDH locks on the axial, tip, and tilt degrees of freedom of the cavity end mirrors. \textbf{a.} Three 785~nm locking beams are phase modulated at distinct, incommensurate frequencies and coupled into separate locking cavities. Their transmitted signals are detected on a common photodiode, independently demodulated, and fed back to the three piezoelectric actuators (PZT) controlling the cavity length, and end mirror tip and tilt. 780~nm "probe" light is coupled into and out of the beam path using dichroics. \textbf{b.} The locking cavities consist of the central cavity (z) used to stabilize the overall CAM length, and two cavities (x and y) displaced from the center cavity by 8 lattice sites to stabilize degeneracy gradients across the array. Initially, the three cavity resonances are not overlapped (inset). As the three feedback loops are engaged, each cavity is brought onto resonance and the total CAM transmission increases to $t_{x}+t_{y}+t_{z}$.}
	\label{fig:degen_lock_fig}
\end{figure*}

Stable operation at degeneracy is essential for continuous operation of the CAM and for ensuring the frequency homogeneity of photons in an imaging or networking scheme. As discussed in Section~\ref{sec:degeneracy}, the dominant sources of non-degeneracy are lens transverse displacements and mirror tip/tilt, both of which produce linear cavity-length gradients across the array. Thus, degeneracy can be maintained by active stabilization of three degrees of freedom: the overall CAM length and mirror tip and tilt.

Experimentally, these degrees of freedom are controlled using (1) a ring piezo mounted on the incoupling mirror to stabilize the overall cavity length, and (2) a piezoelectric tip-tilt mount on the outcoupling mirror to stabilize the transverse degrees of freedom. A fast PDH lock on the central cavity is used to stabilize the overall cavity length ($z$), while two slower PDH locks on cavities displaced by eight lattice sites in $x$ and $y$ stabilize the cavity-length gradients associated with mirror tip and tilt, bringing all cavities into simultaneous resonance (Sup. Fig.~\ref{SI:degenlock}a). We address each cavity with beams that are phase modulated at incommensurate frequencies and read out on a single photodiode. Independent demodulation of the transmitted signal then yields error signals for the $x$, $y$, and $z$ control loops.

Although the CAM is designed for operation at 780~nm, we choose to lock at 785~nm, so that the locking and probe beams can be separated with a dichroic mirror. Experimentally, we find that chromatic aberration shifts the 785~nm cavity stability region by about 1.5~$\upmu$m relative to 780~nm. This shift is substantially smaller than the $\sim11 \upmu$m cavity stability range, allowing the cavities at 780~nm to be stabilized using the 785~nm locks.

\section{Improving cavity finesse}
\label{SI:finesse_opt}
As enumerated in Table~\ref{etab:losses2} of the main text, each optic within the CAM introduces loss which degrades the cavity finesse. At present, these optics are far from optimal and several straightforward modifications, including improved AR coatings, reduced surface roughness, and replacement of the microlens array could increase the cavity finesse beyond 1000.

The largest source of loss comes from the aspheric lenses. Most of this loss arises from non-idealities in the broadband dielectric AR coating, which are exacerbated by the highly curved aspheric surface. To improve the AR coating performance, we are exploring ion beam sputtered (IBS) coatings that are specially tailored to the mode of the CAM. IBS coatings can achieve extremely low optical loss, and new designs project $\sim0.1\%$ per pass. The surface roughness of these aspheres (8-12~nm RMS) also contributes about $0.25\%$ loss per pass. Reducing the surface roughness to 4~nm RMS per side would bring this scattering loss down to $0.02\%$ per pass. 

In addition to the aspheres, the two spherical lenses add an additional $0.07\%$ loss per-pass. For these lenses, we currently use stock Thorlabs substrates with a custom AR-V coating. However, previous work has demonstrated that by using spherical lenses with ultra-low ($<2$\AA) surface roughness, it is possible to achieve single-pass losses as low as 0.008$\%$~\cite{jaffe2021aberrated}. Finally, loss from the microlens array ($0.2\%$ round-trip) can be eliminated by replacing it with an HR coated micromirror array that serves as one of the cavity end mirrors.

Collectively these improvements would reduce the intrinsic round-trip loss of the CAM to approximately $0.59\%$, corresponding to cavity finesses exceeding 1000, cooperativities approaching 100, and readout-optimized photon collection efficiencies greater than $80\%$.

\section{Applications in biological sensing}
\label{SI:biosensing}
\begin{table}[ht!]
    \centering
    \caption{\textbf{Cavity parameters} }
    \vspace{2mm}
    \begin{tabular}{l l l}
        \toprule
        \textbf{ } & \textbf{\makecell{CAM \ \ \ \ \ \ \ }} & \textbf{\makecell{Ref~\cite{needham2024labelfree, kohler2021tracking} \ \ }} \\
        \midrule
        Mode waist   & 1.15~$\upmu$m & 1.1-1.5~$\upmu$m\\
        Number of cavities & 600 & 1\\
        Finesse  & 100-1,000 & 15,000-60,000 \\
        Length & 14~cm & 5-20~$\upmu$m\\
        Field of view \ \ \ \ & 150~$\upmu$m & 2~$\upmu$m\\
        \bottomrule
    \end{tabular}
    \label{etab:paramcomparison}
\end{table}

\begin{table}[ht!]
    \centering
    \caption{\textbf{Proposed CAM geometry} }
    \vspace{2mm}
    \begin{tabular}{l l}
        \toprule
        MLA focal length (mm) & 2 \\
        MLA pitch ($\upmu$m) & 150 \\
        Spherical lens focal length (mm)  & 35 \\
        Aspheric lens focal length (mm) & 1.5 \\
        Magnification & 23 \\
        Cavity spacing at mode waist ($\upmu$m) \ \ \ \ \ & 5\\
        \bottomrule
    \end{tabular}
    \label{etab:CAMgeometry}
\end{table}

Here we consider applying the CAM architecture to Ref.~\cite{needham2024labelfree} and \cite{kohler2021tracking}, which use optical cavities to track the label-free dynamics of solution-phase biomolecules. These cavities are typically integrated into a microfluidic channel and have mode waists of 1.1-1.5~$\upmu$m, cavity lengths of 10-20~$\mu$m, and finesses between 15,000 and 60,000. While such devices provide exceptional sensitivity, they are generally limited to a single sensing volume with restricted spatial tracking range. 

By using a CAM with the parameters suggested in Sup. Table~\ref{etab:paramcomparison} and \ref{etab:CAMgeometry}, the same sensing modality could be extended to create a multi-pixel cavity sensor. With over 500 cavity sensing sites spaced by several microns, the CAM would enable tracking of biomolecular diffusion and transport over distances approaching 100~$\upmu$m. More generally, the large millimeter-scale separation between the intra-cavity aspheres provides sufficient physical access to incorporate interchangeable microfluidic modules within the cavity volume, enabling a high-throughput, sensing platform that can be reused for a variety of biological and chemical sensing applications. Additionally, the shared optical infrastructure of the CAM provides a natural route to suppressing common-mode noise. Cavities located outside of the sensing region could be used to lock and stabilize the array, while resonance shifts of sensing cavities are monitored using techniques such as heterodyne detection. 

The expanded field of view of the CAM is achieved at the expense of reduced cavity finesse relative to state-of-the-art single-cavity sensors. Determining the optimal tradeoff between spatial coverage and sensitivity will require further investigation; nevertheless, the CAM provides a scalable architecture for extending cavity-enhanced biosensing from a single sensing volume to hundreds of spatially resolved sensing sites. 

\section{Alignment procedure}
\label{SI:Alignment}
In the main text, we discuss how the cavity array can be decomposed into two parts: An effective confocal cavity composed of the MLA and mirror, and a perfect double telescope 8f system. For alignment we harness this insight as well as several others. Essentially, our alignment procedure consists of first carefully pre-aligning all beams and reference points, then aligning one half of the full CAM system, then the other half, and finally both halves together. As part of this alignment procedure, we make extensive use of motorized translation stages to move certain optics into/out of the beam path with high repeatability. In particular, for the sub-micron positioning of the spherical lenses to reach degeneracy, we use Thorlabs MAX381 piezo positioning stages. Additionally, for certain alignment steps we make use of Bessel beams generated with off-the-shelf axicons (Thorlabs), which can be used for effectively non-diffractive alignment through a series of multiple optical elements~\cite{chen2024quasiray}.

\begin{figure*}[t!] 
    \includegraphics[width=172mm]{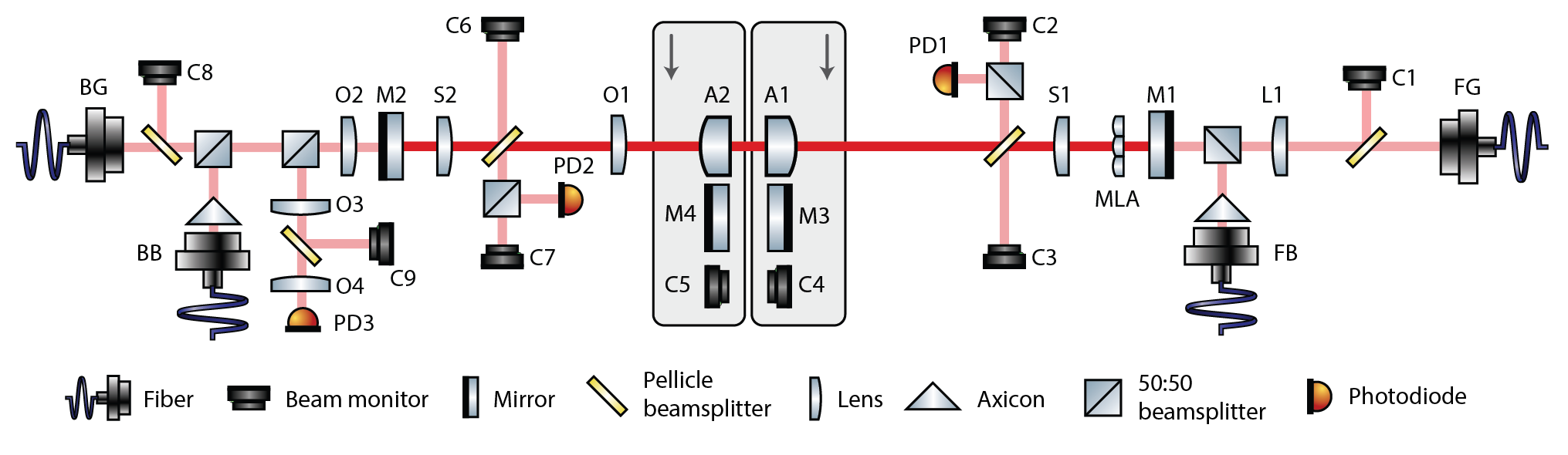}
    \caption{\textbf{Cavity alignment}. Naming scheme for the various optics used during cavity alignment, as referenced in Sup.~\ref{SI:Alignment}. The two large grey rectangles with an arrow in the center are long-throw motorized translation stages to switch out which optic is in the cavity path repeatably. Many cavity optics are placed on individual motorized translation stages (not shown) to enable precise alignment. The axicon optic generates a Bessel-beam which propagates with minimal diffraction for some alignment steps.}
	\label{fig:alignment}
\end{figure*}

In the following alignment procedure, refer to Fig.~\ref{fig:alignment} for optic naming scheme. We assume nearly all optics are not in place at the beginning of the alignment; optics are only inserted when specified. During pre-alignment, we reference the beam position at a number of locations with various cameras; later, we optimize various optics to bring the beam back to this reference. We denote this referencing like so: ``$\textrm{XY}_\textrm{A2}\xrightarrow{FG}\textrm{C3}$'', which means to adjust the x and y positions of the asphere labeled by A2 to bring beam FG back to its reference on camera C3. We denote translation optimization by XY and tip/tilt optimization by TT.

\vspace{2mm}
\textbf{Alignment steps}
\begin{enumerate}

\item Align Front Gaussian (FG) and Back Gaussian (BG) fiber to fiber. Ensure both FG and BG are well-collimated.
\item Align Front Bessel (FB) and Back Bessel (BB).
\item Measure FG on C3, C7, C8; measure BG on C1, C2, C6. Save values as references

\item Align A1
\begin{enumerate}
\item Translate in C5, measure FG on C5
\item Insert A1
\item $\textrm{XY}_\textrm{A1}\xrightarrow{FG}\textrm{C5}$
\item Translate C5 away to let light through, measure retro-reflection of BB on C7
\item $\textrm{TT}_\textrm{A1}\xrightarrow{BB}\textrm{C7}$
\item Iterate
\item Translate A1 out of view
\end{enumerate}

\item Align A2
\begin{enumerate}
\item Repeat above procedure for A2 with corresponding optics
\end{enumerate}

\item Align A2 longitudinal
\begin{enumerate}
\item Translate both A1 and A2 into frame
\item Adjust longitudinal position of A2 so FG is collimated after A1+A2
\end{enumerate}

\item Fine align A1/A2 position and tip/tilt
\begin{enumerate}
\item $\textrm{TT}_\textrm{A1}\xrightarrow{BG}\textrm{C1}$
\item $\textrm{XY}_\textrm{A1}\xrightarrow{BG}\textrm{C2}$
\item $\textrm{TT}_\textrm{A2}\xrightarrow{FG}\textrm{C8}$
\item $\textrm{XY}_\textrm{A2}\xrightarrow{FG}\textrm{C7}$
\item Iterate
\end{enumerate}

\item Align M3
\begin{enumerate}
\item Translate M3 into view
\item $\textrm{TT}_\textrm{M3}\xrightarrow{FG}\textrm{C1}$
\end{enumerate}

\item Align M4
\begin{enumerate}
\item Translate M4 into view
\item $\textrm{TT}_\textrm{M4}\xrightarrow{FG}\textrm{C1}$
\end{enumerate}

\item Align S1 (and S2) XY and tip/tilt
\begin{enumerate}
\item Insert S1
\item $\textrm{XY}_\textrm{S1}\xrightarrow{FG}\textrm{C3}$
\item $\textrm{TT}_\textrm{S1}\xrightarrow{FG}\textrm{C1}$
\item Iterate
\item Repeat procedure for S2
\end{enumerate}

\item Align M1
\begin{enumerate}
\item Insert M1
\item $\textrm{TT}_\textrm{M1}\xrightarrow{FG}\textrm{C1}$
\end{enumerate}

\item Find first half-cavity signal
\begin{enumerate}
\item Translate A1 and M4 into view
\item Look at the beam shape on C2 and C3; before proper alignment, you should see a series of disconnected spots (the multiple passes of the cavity), or a ``comet-like'' shape if the multiple spots are close together. Adjust M1 tip/tilt to make the image symmetric and circular.
\item Measure cavity transmission on PD2
\item Optimize M1 tip/tilt and M4 longitudinal on transmission
\end{enumerate}

\item Align S1 longitudinal
\begin{enumerate}
\item Send large collimated beam (e.g. via FB without axicon) onto the cavity, to couple into many higher-order modes.
\item For arbitrary S1 longitudinal position, the transmission will be a skew-Gaussian, at the optimal telescopic position it will be symmetric; adjust S1 longitudinal to minimize skew, reoptimizing XY as you do so to minimize ``comet-like'' shape on C2/C3.
\end{enumerate}

\item Insert MLA
\begin{enumerate}
\item Remove M1, insert MLA
\item $\textrm{TT}_\textrm{MLA}\xrightarrow{FG}\textrm{C1}$
\item Insert L1; optimize XY and tip/tilt as for S1 (cameras not shown)
\item $\textrm{XY}_\textrm{MLA}\xrightarrow{FG}\textrm{C3}$
\item Insert M1
\item Insert O1; recover cavity signal on PD2
\item Adjust $\textrm{XY}_\textrm{MLA}$ to optimize signal
\end{enumerate}

\item Fine align S1
\begin{enumerate}
\item Use SLM or other method (not shown) to couple into individual modes of the MLA.
\item Measure the degeneracy of the array (adjust A1 longitudinal position to control field of view as needed), as demonstrated in the main text.
\item Adjust $\textrm{XY}_\textrm{S1}$ to minimize linear gradient of the detuning across the array. 
\item Adjust longitudinal position to minimize quadratic phase across the array.
\item (if needed) Rotate the MLA in order to disambiguate the ``true'' degeneracy point
\end{enumerate}

\item Repeat steps 11-13 for the second half-cavity (removing O1)

\item Realize full cavity
\begin{enumerate}
\item Move A1 and A2 into view
\item Insert O2, O3, O4 (these can be pre-aligned if so-wished)
\item Identify cavity transmission on PD3
\item Fine adjust relative asphere distance to maximize transmission for center cavity
\end{enumerate}

\item Fine adjust asphere distance
\begin{enumerate}
\item Take transmission measurements for all cavities while varying the A1-A2 distance
\item Identify stability region and optimal distance, as in Fig.~\ref{fig:fov} of the main text.
\end{enumerate}

\item Repeat step 15 for S2

\item Repeat step 18

\end{enumerate}

\clearpage

\end{document}